\documentclass[aps,amssymb,amsmath,superscriptaddress,twocolumn,nofootinbib]{revtex4-1}
\usepackage{hyperref}
\usepackage{epsfig}
\usepackage{bm}
\usepackage{graphicx}
\usepackage{epstopdf}
\usepackage{physics}
\usepackage{siunitx}
\usepackage{colortbl}
\usepackage[super]{nth}
\usepackage{color}
\usepackage{subfigure}
\usepackage{amsmath}
\usepackage{mathtools}
\hypersetup{
    colorlinks=true,       % false: boxed links; true: colored links
    linkcolor=cyan,          % color of internal links
    citecolor=magenta,        % color of links to bibliography
    filecolor=magenta,      % color of file links
    urlcolor=cyan,           % color of external links
    runcolor=cyan
}

\newcommand{\beqarr}{\begin{eqnarray}}
\newcommand{\eeqarr}{\end{eqnarray}}
\newcommand{\beq}{\begin{equation}}
\newcommand{\eeq}{\end{equation}}

\newcommand{\mc}{\mathcal}

\newcommand{\ignore}[1]{}


\usepackage[printwatermark]{xwatermark}
\usepackage{xcolor}
\usepackage{graphicx}
\usepackage{lipsum}

\begin{document}

\title{Test-driving 
%on the road to quantum supremeness: evaluating performance and avoiding potholes with about 
$1000$ qubits}

\author{Joshua Job}
\affiliation{Department of Physics and Astronomy, University of Southern California, Los Angeles, California 90089, USA}
\affiliation{Center for Quantum Information Science \& Technology, University of Southern California, Los Angeles, California 90089, USA}

\author{Daniel Lidar}
\affiliation{Department of Physics and Astronomy, University of Southern California, Los Angeles, California 90089, USA}
\affiliation{Center for Quantum Information Science \& Technology, University of Southern California, Los Angeles, California 90089, USA}
\affiliation{Department of Electrical Engineering, University of Southern California, Los Angeles, California 90089, USA}
\affiliation{Department of Chemistry, University of Southern California, Los Angeles, California 90089, USA}

\begin{abstract}
Quantum computing is no longer a nascent field. Programmable quantum annealing devices with more that $1000$ qubits are commercially available. How does one know that a putative quantum annealing device is indeed quantum? How should one go about benchmarking its performance and compare it to classical algorithms? How can its performance be improved by error correction? In this contribution to the focus collection on ``What would you do with $1000$ qubits?", we review the work we and others have done in this area, since the first D-Wave quantum annealer with $108$ qubits was made available to us. Some of the lessons we have learned will be useful when other quantum computing platforms reach a similar scale, and practitioners will attempt to demonstrate quantum speedup.
\end{abstract}
\maketitle

\section{Introduction}
As we look forward to the rapid development of new quantum computing devices with hundreds or a few thousand qubits, particularly commercial devices and non-gate-based devices such as quantum annealers, we are faced with a 
challenge. How does one ensure such devices really do what they claim, and aren't effectively classical? How does one evaluate the performance of such a device, what methods should one use to estimate performance on a given metric, and what metrics should one use? How do we do maintenance on the quantum state and ensure we can prevent or correct breakdowns and errors? These questions have to be settled before we can decide where to take our device on a test drive,
and what problems we should use our quantum computing devices to try to solve.

At this time, these new devices and plans for quantum annealing devices and various other quantum computing platforms are no longer the first of their kind. Several generations of programmable quantum annealers from D-Wave Systems have been made available to a small community of researchers, which has worked hard to answer the aforementioned questions. This community began largely groping in the dark, and has over the last five years answered many of the most basic questions, developing techniques to validate quantum annealers, methods to benchmark and estimate performance, and developing methods to suppress errors given the constraints of existing quantum annealers.

We have been fortunate to be members of the aforementioned community, which has given us an opportunity to work with the first several generations of quantum annealers, starting from the first commercially available such device, the $128$-qubit D-Wave One ``Rainier" processor, through two more generations of $512$ and $1152$ qubits, to the current $2048$-qubit D-Wave 2000Q processor.\footnote{A brief history: the Rainier processor ($108$ operational qubits) was the first to be installed at the USC-Lockheed Martin Quantum Computing Center at the USC Information Sciences Institute in 2011. Upgrades to the ``Vesuvius" ($504$ operational qubits) and ``Washington" ($1098$ operational qubits) processors followed in 2013 and 2016, respectively. Google installed ``Vesuvius" ($509$ operational qubits) and ``Washington" ($1097$ operational qubits) processors in the same years at NASA Ames. Los Alamos National Lab installed a ``Washington" processor in 2016. The 2000Q processor is being deployed this year.} As such, rather than answering the question ``what would you do with $1000$ qubits?", in this work we will answer the question ``what have we done with $1000$ qubits?". The discussion will draw mainly from the research we have done on quantum annealers, and we apologize in advance to the many others who have contributed to this enterprise for not doing their work justice. We expect that some of the lessons learned will inform studies of future classes of quantum computing devices with many qubits. Our presentation aims to remain at a fairly high level, without giving a detailed technical account, for which we refer the reader to the original literature cited. 

\section{Quantum Validation Testing}
\label{sec:QVT}

Perhaps the first question one might ask when offered a quantum computational device is whether or not it is, in fact, quantum. In the case of quantum computational devices based on the circuit model and/or gates for quantum computing, the task of validation can be reduced to a Clauser-Horne-Shimony-Holt (CHSH) test between two parts of the device that are treated as black boxes \cite{Reichardt:2013db}. Alternatively, one may opt for quantum process tomography \cite{Chuang:97c,Mohseni:2008ly} or quantum gate set tomography \cite{blume2013robust,Greenbaum:2015aa}, wherein one applies many small computations and measures the results, verifying that they match the predictions of quantum theory. These predictions are available because the quantum computations in question typically involve few qubits and are thus readily implementable \cite{Childs:00,Blume-Kohout:2017aa}. 

However, for other quantum computing paradigms, such as quantum annealing (QA) \cite{kadowaki_quantum_1998,RevModPhys.80.1061} and the broader field inspired by adiabatic quantum computing (AQC) \cite{farhi2001quantum,2002quant.ph.11152K,Kaminsky:2004fk,Albash-Lidar:RMP}, quantum tomography is not currently available for validation. This is for a variety of reasons. The key difference is that gate-based computations are modular: they can be broken into discrete time-local and space-local operations, operating effectively on only one or two qubits at a time, with the others left essentially unaffected, so the only requirement to validate even a long chain of computations is to validate those one- and two-qubit operations on individual qubits and pairs of qubits. For AQC-like platforms, the quantum computation is composed of a continuously time-varying Hamiltonian with many computational operators acting on the system at the same time. They are non-modular in the sense that they cannot be easily broken down into discrete chunks which can be validated separately. Future versions of such platforms may be more flexible and allow for approaches such as quantum tomography, but will still be unable to validate arbitrarily large computations due to the aforementioned nonmodularity of the computation. Meanwhile, partial alternatives such as tunneling spectroscopy have already been explored \cite{Berkley:2013bf}. Of course, in the absence of error correction and fault tolerance neither the gate model nor AQC are guaranteed successful validation.

Nevertheless, certain lessons can be ported over to non-gate-based approaches. One should, as in the circuit model, focus on small problems, with a small number of qubits, and one may hope that by studying such problems applied to many such overlapping sets that one can at least partially validate the operation of the device. From here, two paths for validation become available, depending on whether one can ``open the black box'' and perform measurements during the anneal or use measurements beyond what may be considered ``native'' to the device, or whether one is only able to use the device's output at the end of complete runs for testing. 

\begin{figure}[t]
 \centering
  \includegraphics[width=.9\columnwidth]{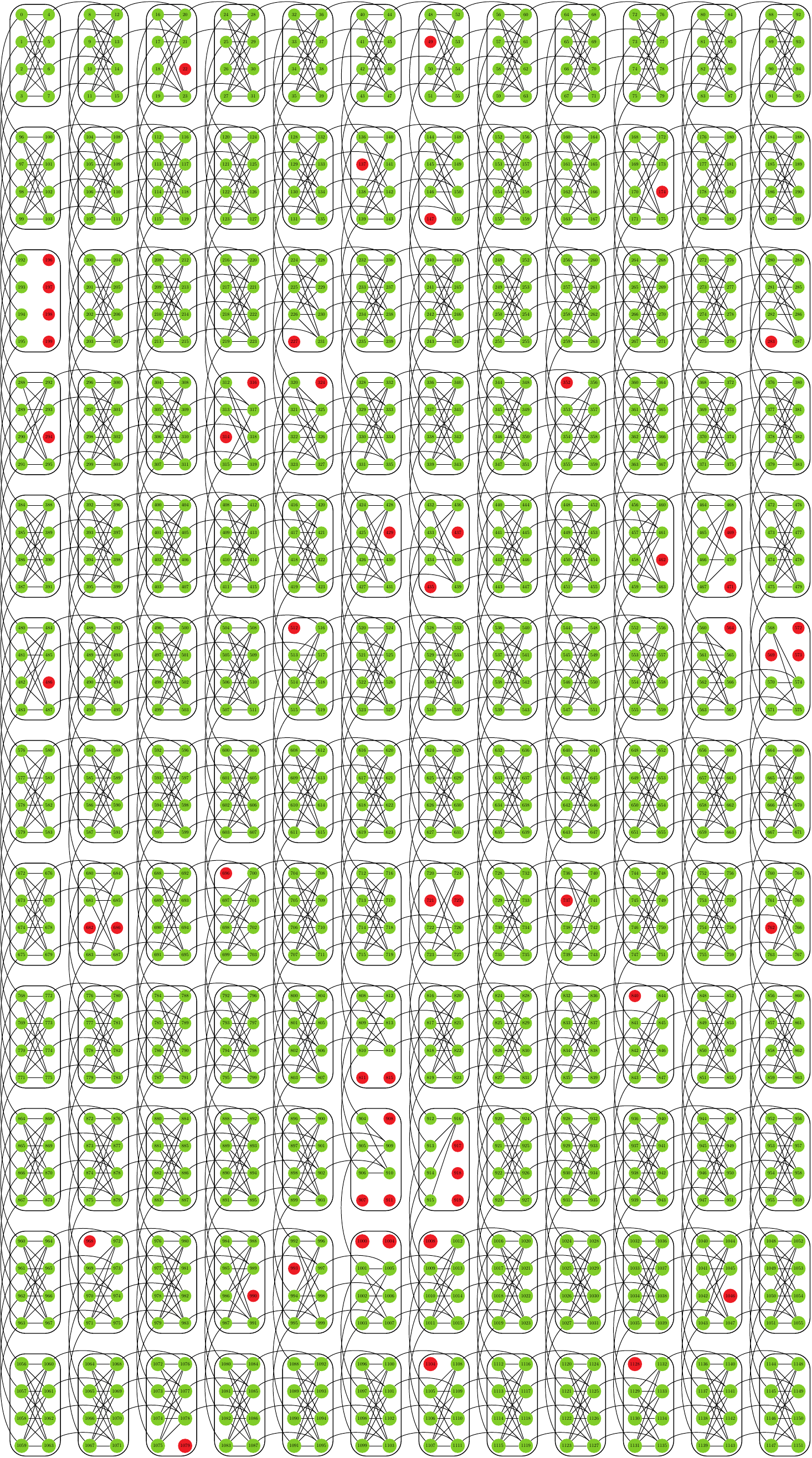}
  \caption{An $1152$ qubit Chimera graph describing the D-Wave Two X processor at the University of Southern California's Information Sciences Institute. Inactive qubits are marked in red, active qubits ($1098$) are marked in green. Black lines denote active couplings (where $J_{ij}$ is programmable to be in the range $[-1,1]$) between qubits.}
  \label{fig:chimera}
\end{figure}

 \begin{figure}[t]
 \centering
  \includegraphics[width=0.9\columnwidth]{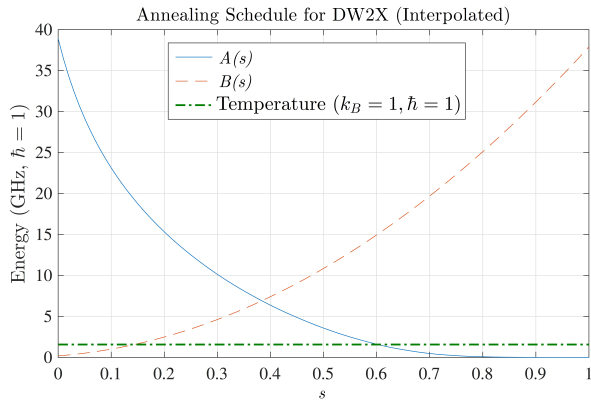}
  \caption{Annealing schedules for the D-Wave Two X processor described in Fig.~\ref{fig:chimera}.}
  \label{fig:schedule}
  \end{figure}

\subsection{Types of validation: proof of quantumness, quantum supremacy, speedup-inferred quantumness, and classical model rejection}
As a case in point, for AQC-style algorithms a system is typically initialized in the ground state of a simple driver Hamiltonian, in most cases a transverse field ``driver Hamiltonian" $H_0=-\sum_{i=1}^N \sigma_i^x$ for $N$ qubits (the $\sigma$'s are Pauli operators), and then the Hamiltonian is slowly modified into the ``problem Hamiltonian'' $H_1$ via the transformation 
\begin{equation}
	H(s) = A(s)H_0 + B(s)H_1\ ,
	\label{eq:Heq}
\end{equation}
where $A(s)$ and $B(s)$ are monotonically decreasing and increasing functions of the dimensionless time $s=t/t_f$, respectively, where $t\in [0,t_f]$ and $t_f$ denotes the annealing time.
Here $H_1$ encodes the problem via programmable parameters $\{h_i\}$ (``local fields" or ``biases") and $\{J_{ij}\}$ (``couplers"): 
\begin{equation}
H_1 = \sum_{i=1}^N h_i \sigma_i^z + \sum_{i<j}^N J_{ij} \sigma_i^z \sigma_j^z\ .
\label{eq:H_1}
\end{equation}
For the D-Wave quantum annealers the $h_i$ and $J_{ij}$ terms are programmable with values bounded between $[-2,2]$ and $[-1,1]$, respectively. The form in Eq.~\eqref{eq:Heq} is the general form for an Ising model quantum annealer, and as written every qubit in the final Hamiltonian is connected to every other qubit. In reality (for example the D-Wave architecture), full connectivity is difficult to achieve, and as such there may be additional restrictions on the $J_{ij}$, such that they can be nonzero iff the nodes $i$ and $j$ are connected on the hardware graph of the device. An example of the ``Chimera" hardware connectivity graph of a D-Wave Two X (DW2X) processor is shown in Fig.~\ref{fig:chimera}, and its ``annealing schedule'', the $A(s)$ and $B(s)$ curves, are shown in Fig.~\ref{fig:schedule}.

In validating quantum annealers, one seeks to create an assignment to the $h$'s and $J$'s such that one can take some measurements which will conclusively demonstrate, for instance, quantum entanglement, in what might be called an experimental ``\emph{proof of quantumness}''. 

A somewhat weaker and indirect type of validation is provided by ``\emph{quantum supremacy}" experiments \cite{Preskill:2012aa},
since they have the potential for complexity theoretic guarantees.\footnote{The term ``supremacy" has generated considerable controversy \cite{Qsupremacy-debate}. While we would prefer the adoption of an alternative such as ``hegemony" or ``supremeness", we recognize that ``supremacy" is likely here to stay due to its current widespread usage.} 
 More specifically, quantum supremacy is a scenario where (part of) the polynomial hierarchy of complexity theory collapses if the quantum result could be replicated classically without slowdown \cite{Aaronson:2016aa,Bremner:2016aa,FarhiHarrow-QAOA,Gao:2017aa,Boixo:2016aa,Fefferman:2017ab}. While weaker than a direct proof of quantumness, a demonstration of quantum supremacy would be considered strong evidence for quantum computational power of a device, which may be considered inherently more interesting than a direct demonstration of, e.g., entanglement. 

``\emph{Speedup-inferred quantumness}" is a related type of indirect validation based on a demonstration of quantum speedup \cite{speedup} over the best classical solvers known for a task, which is often considered the holy grail of quantum information processing. Unlike quantum supremacy tests, speedup-inferred quantumness tests do not have complexity theoretic guarantees (an example in the circuit model would be Shor's algorithm \cite{Shor:97}). It appears that an unqualified quantum speedup would necessarily have to invoke quantum properties, and this might happen even if these properties remain poorly understood or characterized. Thus a certificate of quantumness might be assigned even in the absence of a direct demonstration of quantum properties such as entanglement. It should be recognized that this carries a certain element of risk. For example, suppose a new \emph{classical} optimization is discovered that outperforms all other classical and quantum optimization algorithms known to date (this is in fact what happened recently in a tug-of-war between quantum and classical optimization for the Max E3LIN2 problem \cite{Farhi:2014aa}). This algorithm could be deceptively marketed as a quantum algorithm providing speedup-inferred quantumness by a shrewd company claiming to own quantum computers, that provides black-box access only to run the new optimization algorithm. Thus any claim of speedup-inferred quantumness should always be treated with a healthy degree of skepticism as related to its quantum underpinnings, until actual evidence of quantum effects driving the algorithm is presented.

If none of prior three types (proof of quantumness, quantum supremacy, speedup-inferred quantumness) of validation are attainable, one may alternatively seek to show that on sufficiently small scale problems the results are only readily reproducible using a truly quantum model of the device, and cannot be replicated qualitatively using any existing classical model, in what might be called ``\emph{classical model rejection}''. This type of validation experiment does not provide a certificate of quantumness, since one can always invent a new and better classical model. Instead, one can only hope to exclude all ``physically reasonable'' classical models for the device. Moreover, classical model rejection can only be performed as long as it is feasible to carry out quantum model simulations, which limits system sizes to about $20$ qubits for master equation type models, using the quantum trajectories method \cite{Yip:2017}. Extrapolations to larger sizes are, as always, risky in the absence of fault tolerance guarantees.

One caveat regarding ``proof of quantumness" experiments is noteworthy. While demonstrations of entanglement can be considered ``proof of quantumness", they often require additional physical resources and measurement possibilities beyond those that may natively be embedded in a (commercial) quantum computational device or that are strictly required to implement the core algorithm, and thus may be impossible on certain platforms. Additionally, in practice, certain assumptions may be made in a ``proof of quantumness'' experiment which, when relaxed, render it effectively a ``classical model rejection'' experiment; we shall shortly see an example of this with the D-Wave quantum annealers.

\subsection{Experimental implementations of quantum validation tests}
\label{sec:QVT-expt}

The primary ``proof of quantumness'' experiment for quantum annealers was performed in Ref.~\cite{DWave-entanglement}, using an entanglement test on the D-Wave Two (DW2) generation of processors. Briefly, the work used quantum tunneling spectroscopy \cite{Berkley:2013bf} to estimate the populations of the first and second excited states of a combined probe-system Hamiltonian. They also measured the energy spectrum and found it to be consistent with the Hamiltonian the device was designed to implement, which provided a justification for the assumption that the measured populations were those of the energy eigenstates of the Hamiltonian.
This allowed for a reconstruction of the density matrix under the assumption that it is diagonal in the energy eigenbasis, enabling a computation of the negativity \cite{Vidal:02a} for all possible bipartitions of the system, the geometric mean of which was taken as a measure of the entanglement of the system. As it was found to be nonzero, the system is entangled. Further, by exploiting the theory of entanglement witnesses~\cite{Spedalieri:2012fk}, 
Ref.~\cite{DWave-entanglement} was able to show that even if the diagonality assumption is relaxed, the entanglement remains. This was used to conclude that the DW2 system tested displays entanglement at least on the scale of a single $8$-qubit unit cell.

It was noted in Ref.~\cite{Albash:2015pd} that these tests depended on the assumption that the device was well-described by Eq.~\eqref{eq:Heq} for an appropriate (programmed) choice of local fields and couplers for which the ground state is entangled, and that this assumption is not directly demonstrable by the experiments in \cite{DWave-entanglement}. Without that assumption, one must revert to a ``classical model rejection'' experiment in which one compares results of direct quantum simulations of the device and available classical alternatives to demonstrate that only the quantum model is consistent with the experimental observations. Ref.~\cite{Albash:2015pd} provides a detailed description of the experiments, but for our purposes the key takeaway is that only the quantum adiabatic master equation \cite{aqcME} can reproduce the output distribution from experiments, validating the approach in Ref.~\cite{DWave-entanglement}.

\begin{figure}[t] 
\centering
 \includegraphics[height=1.75in]{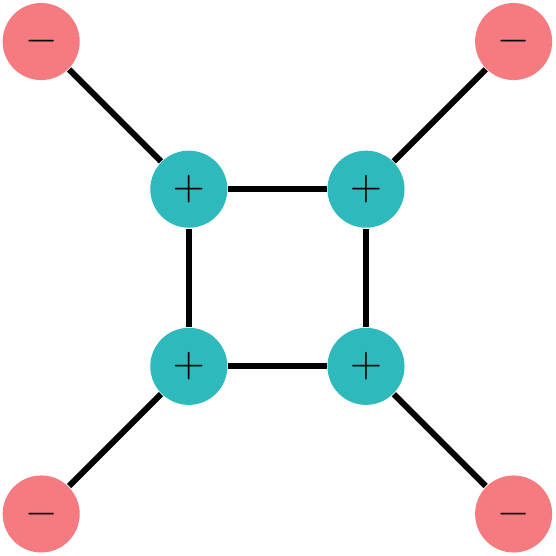} 
\caption{The eight-spin Ising quantum signature Hamiltonian introduced in Ref.~\cite{q-sig}. The inner ``core" spins (green circles) have local fields $h_i = +1$ while the outer spins (red circles) have $h_i = -1$.  All  couplings are ferromagnetic: $J_{ij} = 1$ (black lines).}
\label{fig:signaturehamiltonian}
\end{figure}

Another branch of validation experiments of the classical model rejection type are the so-called ``quantum signature'' Hamiltonians and the consistency tests derived therefrom, introduced in Ref.~\cite{q-sig}, critiqued in Ref.~\cite{Smolin}, and further explored in Refs.~\cite{comment-SS,q-sig2,SSSV-comment}. Unlike the aforementioned entanglement tests, these experiments do not require access to the system during the annealing process, and are appropriate for cases in which the quantum device is a ``black box'' in which one can only control the inputs and measure the outputs. An example of a quantum signature Hamiltonian is shown in Fig.~\ref{fig:signaturehamiltonian}. These Hamiltonians take the form of a ring of tightly bound qubits each connected to a single outer qubit. The resulting Hamiltonian has the property that there is a large ($2^{N/2}$-dimensional) degenerate subspace of ground state configurations corresponding to arbitrary assignments to the outer qubits where all qubits in the inner ring are in the state $0$ (forming a ``cluster" connected by single spin flips applied to the outer ring). There is one additional ground state corresponding to flipping all the inner qubits to the state $1$, dubbed the ``isolated" state, since for a signature Hamiltonian with $2N$ qubits it is at least $N$ spin flips away from all other ground states. Thermal algorithms such as classical simulated annealing (SA) \cite{kirkpatrick_optimization_1983} will be weighted toward the isolated state, such that it will have the highest probability of occurrence of any ground state configuration, whereas an adiabatic quantum evolution will find the isolated state to be suppressed relative to the cluster states. Extensive simulations and experiments on a $108$ qubit D-Wave One (DW1) ``Rainier" processor matched qualitatively with the adiabatic master equation across all the parameters and statistics of the output distributions tested, though noise due to cross-talk made it very difficult to find quantitative agreement; at the same time all existing classical models failed to qualitatively match the DW1 in at least one of the tests \cite{q-sig2}.

A different approach to classical model rejection was taken in Ref.~\cite{q108}, which used random $J_{ij}=\pm 1$ instances of the Ising Hamiltonian in Eq.~\eqref{eq:H_1} to test the hypothesis that quantum annealing correlates well with two classical models: SA and classical spin dynamics \cite{Smolin} (also known as the Landau-Lifshitz-Gilbert model). The hypothesis was tested using the same DW1 processor. This work showed that these two classical models failed to correlate with the results for the distribution of ground state probabilities generated by the DW1 device, while the DW1 correlated very well with simulated quantum annealing (SQA), implemented using quantum Monte Carlo \cite{sqa1}. This was taken as evidence for quantum annealing on the scale of more than $100$ qubits, thus generalizing the conclusion of the earlier result \cite{q-sig} based on the $8$-qubit ``gadget" shown in Fig.~\ref{fig:signaturehamiltonian}. Shortly thereafter a new semiclassical Spin-Vector Monte Carlo (SVMC) model was introduced, also known as SSSV, the author initials of Ref.~\cite{SSSV}.\footnote{Both the spin-dynamics and SVMC models can be derived in a strong coupling limit from the anisotropic Langevin equation, starting from Keldysh field theory \cite{Crowley:2016aa}.} 
In this model spins are treated as $O(2)$ rotors (effectively as single qubits), evolved according to the annealing schedule given in Eq.~\eqref{eq:Heq}, with Monte Carlo angle updates. The SVMC model correlated well with both the DW1 and SQA data, suggesting that although the DW1 device's performance is consistent with quantum annealing, it operated in a temperature regime where, for most random Ising spin glass instances, a quantum annealer may have an effective semi-classical description. This conclusion was challenged in Ref.~\cite{Albash:2014if}, which considered the excited state distribution rather than just the ground state distribution over random $J_{ij}=\pm 1$ Ising instances, as well as the ground state degeneracy. This work presented evidence that for these new measures neither SQA nor SVMC, which are classically efficient algorithms, correlated well with the DW1 experiments. The close correlation SQA and the SVMC model was explained by showing that the SVMC model represents a semiclassical limit of the spin-coherent states path integral, which forms the foundation for the derivation of the SQA algorithm.

The intense debate that arose around the original classical model rejection tests presented in Refs.~\cite{q-sig,q108}, in particular the critique presented in Refs.~\cite{Smolin,SSSV}, illustrates the risks associated with such tests---risks that materialize whenever a sufficiently clever new classical model is found that agrees with (some of) the data---as well as the fruitfulness of the classical model rejection approach, which can lead to a healthy updating and sharpening of models and assumptions.

Black box classical model rejection tests such as the quantum signature Hamiltonians provide the basis for the testing of new putative quantum devices for which available controls and potential measurements are limited, and ultimately even the best experiments that seek to prove entanglement will depend on a series of such experiments to demonstrate that only quantum models can reproduce the experimental data from the device. Quantum supremacy tests are a type of limiting case of this, in which one can prove that should any classical device be able to produce a particular output distribution in polynomial time then the computational complexity hierarchy will at least partially collapse. Since this is not expected to occur, building a device which can produce said distribution efficiently will then immediately rule out all classical models for the device \cite{Aaronson:2016aa}.

Another kind of black box classical model rejection test is based on the phenomenon of quantum tunneling, whereby a quantum state has sizable probability on either side of an energy barrier which the system could not move through classically, or at least will only be able to do so with reasonable probability at high temperature. The first quantum annealing experiments involving tunable
tunneling were carried out using the disordered ferromagnet LiHo$_x$Y$_{1-x}$F$_4$ in a transverse magnetic field \cite{Brooke1999,brooke_tunable_2001}, and served as inspiration for the design of programmable superconducting flux-qubit based quantum annealers. These experiments indicated that quantum annealing hastens convergence to the optimum state via tunneling, compared to simple thermal hopping models. 
The first programmable quantum annealer experiment was reported in Ref.~\cite{DWave}, in which it was demonstrated that an $8$-qubit quantum annealing device was able to reproduce the domain wall tunneling predictions of quantum theory for a chain of superconducting flux qubits by modifying the time during the annealing process at which a local field is abruptly applied to the qubits. This contradicted the temperature dependence predictions of a classical thermal hopping model, thus serving as a classical model rejection experiment.

\begin{figure}[t] 
\centering
 \includegraphics[height=1.75in]{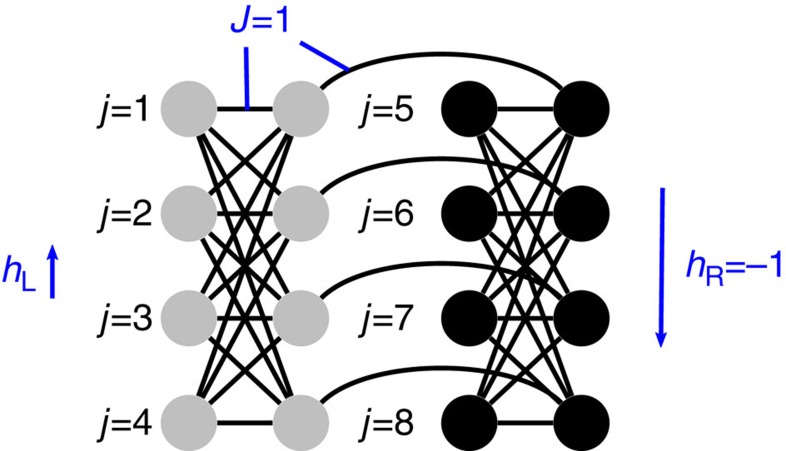} 
\caption{The $16$-spin Ising Hamiltonian composed of two $K_{4,4}$ unit cells introduced in Ref.~\cite{Boixo:2014yu}. All couplings are set to $J=1$, all qubits in the left unit cell have a local field $0<h_L<0.5$ applied to them while all spins in the left unit cell have $h_R=1$ applied to them. Two local minima form, one with the cells internally aligned but in opposite states from each other (a local minimum) and the other with all states aligned with $h_R$ (the global minimum). By tightly binding each unit cell, they effectively act like single large spins.}
\label{fig:tunnelingprobe}
\end{figure}

More recently, Ref.~\cite{Boixo:2014yu} reported on a specially designed tunneling probe Hamiltonian for quantum annealing, illustrated in \ref{fig:tunnelingprobe}. The probe uses two unit cells of the D-Wave Chimera graph, binding each one together tightly so they each act like a single effective spin, or cluster. Opposite magnetic fields are applied to each unit cell, one weak and one one strong, so that the spins in the ``strong" cluster align before the spins in the ``weak" cluster.
Initially, there is only a single minimum. A second minimum develops over the course of the anneal, and eventually becomes the global minimum of the final Ising Hamiltonian.
The only way to reach the global minimum is to overcome an energy barrier whose strength increases as the anneal progresses, a classic example of tunneling. Using the non-interacting blip approximation (NIBA) it was shown in Ref.~\cite{Boixo:2014yu} that the system effectively acts like a two-level system even in the open-system setting with a strongly coupled bath. NIBA-based predictions without free parameters 
for tests at different values of $h_L$ and different temperatures demonstrated very good agreement with experiments involving a DW2 device, and were not reproducible using classical models for the device such as SVMC \cite{SSSV}. A variant of this experiment was reported on in Ref.~\cite{PhysRevX.6.031015}, which introduced a new class of problem instances which couples the weak-strong clusters of the tunneling probe as sub-blocks of the Hamiltonian. This work can be interpreted as an attempt to go from classical model rejection to speedup-inferred quantumness, as it claimed a large tunneling-induced constant-factor speedup over classical simulated annealing and simulated quantum annealing for a DW2X device. However, this claim was critiqued in Ref.~\cite{2016arXiv160401746M} on the basis of a comparison to classical algorithms with better performance. Moreover, as we discuss below, speedup-inferred quantumness requires a demonstration of an optimal annealing time \cite{speedup}, which was absent in the results reported in Ref.~\cite{PhysRevX.6.031015}.

Validating non-gate based quantum devices will continue to be a challenge as new such systems come online, but applying combinations of the techniques discussed above, from the construction of quantum signature Hamiltonians and tunneling probes to (in)direct proofs of entanglement via entanglement witnesses and direct computation of entanglement, should allow one to boost confidence that the system obeys the predictions of quantum theory \emph{over small scales}.
The challenge remains to extend these techniques so that they are able to demonstrate conclusively that a device with hundreds or thousands of qubits displays coherence and long-range entanglement. 
Due to decoherence this presents a challenge for gate-based quantum devices as well even at a smaller scale \cite{PhysRevLett.106.130506,Bohnet:2016aa}, and speedup-inferred quantumness tests may prove to be simpler to execute than direct quantumness tests even in the gate-model setting. 
%\JJ{This remains a challenge for gate-based quantum devices as well, and it may well be that direct quantumness tests will have to be replaced by an appeal to speedup-inferred quantumness. JJ: This doesn't seem true; once you've validated the one and two qubit operations, you can just drive the system to $\ket{+}^N+\ket{-}^N$ and measure in the X basis and Z basis and you're done, right?}

Once one has validated that the device works in approximately the manner one would expect from the instruction manual, one can then turn to the question: ``Toward what practical purpose may this device be put?". The choice of appropriate problems in this domain is a complex issue that we cannot address here. Indeed, before we can answer that question, we must first focus on an operational question: ``How does one go about comparing performance in a specified problem domain between a verified quantum computing device and existing classical strategies?". We discuss this next.

\section{Benchmarking}
\label{section:benchmarking}

Assume we have at our disposal a device verified to be quantum, at least provisionally on the small scales covered by classical model rejection, and we would like to compare its performance to competing classical solvers. This is the task we refer to here as benchmarking, which belongs more generally to the field of experimental algorithmics \cite{McGeoch:book}. Specifically, consider the problem of estimating the value of some function of merit (or ``reward'') $R$ from the output of a given solver (e.g., our quantum device or some classical algorithm) for a given problem family $\mc{P} = \{P\}$. Each problem instance $P$ is parametrized by some parameters $\theta$. In the case of quantum annealers, particularly studies of the D-Wave devices thus far, the goal has generally been to find the ground state of Ising Hamiltonians as defined in Eq.~\eqref{eq:H_1}. In that context, typically the reward is taken to be the negative of the time to solution (TTS), defined as $\text{TTS}=t_f \log(1-p_d)/\log(1-p)$ for a probability $p$ of finding the ground state at least once with desired probability $p_d$ (typically $0.99$), and annealing time $t_f$.\footnote{The probability of not finding the ground state even once after $k$ independent runs of duration $t_f$ each is $(1-p)^k$, so the probability of finding it at least once is $1-(1-p)^k$, which we set equal to $p_d$. Solving for $k$ and substituting into $\text{TTS}=t_f k$ gives the TTS formula. See, e.g., Ref.~\cite{speedup} for a more detailed derivation.} In the language above, $R=-\text{TTS}$ (one would like to minimize the TTS), and the problem is parametrized by $\theta=\{h_i,J_{ij}\}$. Many similar metrics have been proposed, such as time-to-epsilon and time-to-target \cite{King:2015cs}, which amount to mild generalizations of TTS. A more elaborate notion of cost, based on optimal stopping theory, has also been considered and shown to recover the previous metrics as special cases \cite{Vinci:2016tg}. We shall return to this below.

\subsection{Solvers for comparison}
The choice of classical solvers against which to compare the quantum device involves a few considerations. It is important to perform an apples-to-apples comparison, in that if the device is probabilistic, it would be misleading to measure its performance against a deterministic algorithm \cite{McGeoch,q108,MAX2SAT}. For a quantum annealer, a performance comparison to known heuristic algorithms for sampling low-energy states from Ising models is natural, such as SA \cite{kirkpatrick_optimization_1983,Isakov:2015ao}, parallel tempering (PT) \cite{Geyer:91,Earl:2005pd,katzgraber:06a}, and the Hamze-Freitas-Selby (HFS) algorithm (which searches all states on nodes that make up induces trees or small-treewidth subgraphs of the Ising model's connectivity graph) \cite{hamze:04,Selby:2014tx}. One might also compare to approximations of QA itself, in particular simulated quantum annealing (SQA) \cite{sqa1,Heim:2014jf,Crosson:2016fk}, or the SVMC algorithm \cite{SSSV}. All of these can be said to be ``solvers'' for the Ising problem on QA. But, to determine if the quantum device is truly useful in practice, it must also be compared to the \emph{best} algorithm for solving the \emph{original} (typically non-Ising problem) task. For example, when solving the graph isomorphism problem \cite{Zick:2015aa}, job-shop scheduling \cite{Venturelli:2015pi}, operational planning \cite{Rieffel:2015aa}, or portfolio optimization \cite{Rosenberg:2015}, the original problem must first be mapped into an Ising problem \cite{2013arXiv1302.5843L} and then embedded using the existing hardware connectivity graph \cite{Choi1,Choi2,klymko_adiabatic_2012}; the performance of the quantum device must be compared to the best algorithm for solving the original problem, and the mapping plus embedding steps can severely reduce performance. Note also that determining what the truly optimal classical algorithm is can be a daunting, or even impossible, challenge. In many cases one settles for an educated guess: the standard and/or currently best known algorithm(s). Finally, it is important to remember that any tests run on a quantum device that does not enjoy a fault tolerance guarantee cannot be reliably extrapolated to arbitrarily large sizes. I.e., in the absence of such a guarantee, a finite-size device provides evidence of what can be expected at larger sizes only provided that quantities such as the device temperature, coupling to the environment, and calibration and accuracy errors, can be appropriately scaled down. With this in mind, let us turn to a discussion of much of the benchmarking work done so far and some of the considerations that go into using large, noisy quantum devices.

\subsection{The state of benchmarking}

The first comprehensive study benchmarking QA devices was Ref.~\cite{q108}, using a $108$ qubit DW1 processor. This article introduced many of the concepts used in later studies in the field, including the above definition of the TTS. It focused on the performance on the set of random Ising problems with binary $\pm1$ local fields and couplings, and introduced the use of SA and SQA as important comparison algorithms. It also noted the importance of comparing against parallelized versions of classical algorithms, as quantum annealers such as the D-Wave device consume linearly more computational hardware with increasing problem size, and in many cases SA and SQA can be effectively parallelized in much the same way. 

Another significant contribution of Ref.~\cite{q108} was the use of ``gauge averaging'' in benchmarking, a technique that was introduced in Ref.~\cite{q-sig} (where it was called ``spin inversions") and which has become so universal that it is now included natively in the D-Wave API for their processors, and which points toward a more general consideration for noisy quantum devices in the absence of quantum error correction. 
The need for gauge averaging arises from the observation that in QA, one may have per-qubit or per-edge random and systematic biases from stray fields or interactions.
In such cases, performance may be dramatically impacted by the choice of mapping from a logical Hamiltonian as defined in Eq.~\eqref{eq:H_1} to a physically implemented computation.
In essence, a gauge transformation corresponds to swapping which physical spin state corresponds to a computational $0$ or $1$. In an ideal annealer, this transformation commutes with (i.e., is a symmetry of) the total Hamiltonian and so has no dynamical effect. However in the presence of noise, this symmetry is broken and the choice of gauge does make a difference, and indeed was found to have a significant effect on the performance of the DW1 quantum annealer, to such a degree that the device did not even correlate with itself if one compares one gauge to another, or even one gauge with itself when run later (most likely the result of slow drift $1/f$ noise resulting in the effect that each time the annealer is programmed, a small random error term is added to the Hamiltonian). However, when results for the same Hamiltonian were averaged across many gauges, the DW1 processor correlated quite well with itself \cite{q108}. Since then, applying many gauge transformations to the same Hamiltonian and averaging the results has become a standard practice in the QA community, and the idea behind it has been steadily generalized since then to include sampling over every known potentially broken symmetry of the Hamiltonian.

For example, if one is solving a fully connected Ising problem, the Hamiltonian has a permutation symmetry. Since every logical spin has an interaction with every other, one can relabel which spin is which without changing anything about the logical problem. However, when one goes to implement such a problem on an actual quantum annealer with limited connectivity, such as the DW2, one has to perform a minor embedding in which each logical spin is mapped to a chain of spins on the physical device \cite{Choi1,Choi2}. Those physical spins may have local field biases which vary from chain to chain, and thus the distribution over logical states will depend, in part, on the assignment of the logical spin variables to the physical chains, as shown in Ref.~\cite{Venturelli:2014nx}. This work was the first case study of both minor embedding of fully connected problems as well as permutation embeddings for such problems, and demonstrated the importance of optimizing the strength of the coupling in minor embedding applications, a topic which is discussed in more detail in Section~\ref{sec:err-corr}. 

Ref.~\cite{MAX2SAT} demonstrated evidence for the easy-hard-easy phase transition for Max 2-SAT problems (wherein one wishes to find the maximal number of simultaneously satisfiable two-variable Boolean clauses over a set of variables from some ensemble of clauses) near a clause density of one, on the $108$-qubit DW1 processor. It performed a rudimentary benchmarking comparison between the DW1 and an exact Max 2-SAT solver (akmaxsat) (see also Ref.~\cite{McGeoch}), and noted that there was no correlation between the two solvers over randomly selected instances of Max 2-SAT. This work also introduced the important idea of bootstrapping into the QA community, variants of which (such as the Bayesian bootstrap \cite{rubin1981bayesian}) formed the backbone of error analyses for later studies, as a nonparametric method for approximating the distribution over the problem space and over the aforementioned broken computational symmetries.

%The main focus of Ref.~\cite{q108} was in fact not benchmarking, but quantum validation testing, as described in Sec.~\ref{sec:QVT-expt}. 
A decisive step forward was taken 
%later that same year, 
in Ref.~\cite{speedup}, which introduced the notion of different quantum speedup categories. Of particular interest in the benchmarking context are \emph{potential quantum speedup}, defined as a speedup compared to a specific classical algorithm or a set of classical algorithms (e.g., simulation of the time evolution of a quantum system implemented on a quantum computer as compared to directly solving the Schr\"{o}dinger equation on a classical computer), a \emph{limited quantum speedup}, defined as a speedup against algorithms that may be said to be analogous to the quantum solver (e.g., SA or SQA compared with a quantum annealer), and an unqualified \emph{quantum speedup}, defined as a speedup against the best available algorithms for solving the problem (e.g., Shor's algorithm for factoring). A crucial observation made in Ref.~\cite{speedup} was that unless an optimal annealing time can be explicitly demonstrated (i.e., an observed minimum in the TTS as a function of the annealing time), a scaling analysis performed over a finite range of problem sizes cannot be trusted to reveal any type of quantum speedup. The reason is that an annealing time $t_f$ that is too large (suboptimal) can artificially inflate the TTS at small problem sizes, thus leading to artificially shallow scaling, and potentially to a false conclusion that (some type of) quantum speedup is present. These notions were then applied to random Ising instances with a wide range of integer couplings (up to $\lvert J_{ij}\rvert=7$, which is renormalized to $[-1,1]$ when submitted to the processor), and tested on a DW2 device with up to $503$ qubits. The analysis of the scaling of TTS with system size mostly demonstrated a disadvantage for the DW2 over SA, but an advantage for the DW2 over SA on lower hardness percentiles of the problem distribution (i.e., the easier problems). However, due to the aforementioned issue with suboptimal annealing times, this was not taken as evidence of any type of quantum speedup. Very recently evidence of a limited quantum speedup with optimal annealing times was reported \cite{Albash:2017aa}, as we discuss below.

An interesting critique of the scaling results presented in Ref.~\cite{speedup} was made in Ref.~\cite{2014Katzgraber}, which argued that random Ising instances restricted to the Chimera graph are ``too easy", essentially since their phase space exhibits only a zero-temperature transition. This would imply that classical thermal algorithms such as SA see a simple energy landscape with a single global minimum throughout the entire anneal (except perhaps at the very end as the simulation temperature is lowered to near zero), instead of the usual glassy landscape with many local traps associated with hard optimization problems. This work highlighted the importance of a careful design of benchmark problems, to ensure that classical solvers would not find them trivial. Of course, it should be stressed that quantum speedup is always relative, and it can be observed even when efficient (polynomial-time) classical algorithms exist, as in, e.g., the solution of linear systems of equations \cite{PhysRevLett.103.150502}. In light of this one may interpret the message of Ref.~\cite{2014Katzgraber} to mean that a quantum speedup might not be \emph{detectable} over a finite range of problem sizes if the problem is classically easy, since the difference between the quantum and classical scaling is too subtle to be statistically significant.

An attempt to address the critique that random Ising instances are too easy was made in Ref.~\cite{Hen:2015rt}, which introduced a new class of ``planted solution" instances (see also the followup study Ref.~\cite{King:2015zr}, though neither study demonstrated a non-zero critical temperature). The problem Hamiltonian $H_1$ [Eq.~\eqref{eq:H_1}] is constructed out of a sum of frustrated small-loop Hamiltonians, each one designed so its ground state is a chosen bit-string. In so doing, the total Hamiltonian is guaranteed to have as one of its ground states the chosen bit-string, dubbed a ``planted solution''. Knowing a solution in advance is an important advantage of this problem class over the classes tested before, for which solutions could only be found either heuristically or by directly solving the Ising problem at a cost which is generally exponential with the system size (in particular, scaling like $2^{4L}$ for an $L\times L$ unit cell problem on the Chimera graph), which is prohibitive for systems much larger than those tested in previous studies. By knowing a solution in advance, the ground state energy can be computed instantly, and any further global optima can be recognized immediately, providing a sound basis for TTS comparisons for problems that may turn out to be too large for brute force search. This study was also one of the first (following \cite{selby:13b}) to compare against the Hamze-Freitas-Selby (HFS) algorithm \cite{hamze:04,Selby:2014tx}, which has been considered ever since to be the ``algorithm to beat'' thanks to its superior scaling and direct exploitation of the large treewidth induced subgraphs possible in the Chimera graph. It was found in Ref.~\cite{Hen:2015rt} that the DW2 had flatter scaling than all algorithms that had been tested up to that point (SA, SQA, SSV) over virtually the entire range of frustrated loop-to-spin density. In the comparison to the HFS algorithm it was found that the latter was able to achieve superior scaling compared to the DW2 for all but the easiest and largest loop densities. These results invited the possibility of a limited quantum speedup, but due to the lack of an optimal annealing time, this could again not be demonstrated. Moreover, Ref.~\cite{Hen:2015rt}  provided a proof (under the assumption that the TTS increases monotonically with the annealing time) that without an optimal annealing time, one could never conclusively demonstrate even a limited quantum speedup. 

Before we turn to a discussion of the evidence for a limited quantum speedup, we first briefly discuss alternatives to the TTS as a performance measure. One such alternative is the time-to-target (TTT), i.e., the total time required by a solver
to reach the target energy at least once with a desired probability, assuming each run takes a fixed time \cite{King:2015cs}. This reduces to the TTS if the target is the ground state. A unified approach that includes a variety of other measures was presented in Ref.~\cite{Vinci:2016tg}, drawing upon optimal stopping theory, specifically the so-called ``house-selling" problem \cite{Ferguson:book}. Within this framework one answers the question of how long, \emph{given a particular cost for each sample drawn from a solver}, one should sample in order to maximize one's reward, analogously to the decision problem about when to sell one's house given that bids accrue over time but that waiting longer carries a higher monetary cost. This allows the TTS and TTT, among other measures, to be shown to be specific choices of the cost and reward functions. The optimal stopping framework also paves the way for a more detailed comparison between quantum and classical approaches and the tradeoffs of each, as by altering the cost per sample one can see the impact of the distribution over states (rather than just the ground state) for the various solvers. Optimal stopping is appropriate for applications where finding the minimum energy is not strictly the most important consideration for the application, such as many machine learning contexts and even various business-origin optimization problems. In those cases, there is a tradeoff between the cost to perform the computation and the benefit from receiving a result. Tests were performed demonstrating the optimal stopping approach with a DW2X device (with $1098$ qubits) on frustrated loop problems much like those in Ref.~\cite{Hen:2015rt}, demonstrating identical scaling (modulo concerns about the lack of an optimal annealing time) to the HFS algorithm at multiple values of the cost to draw a sample, an improvement over the DW2. However, these results could still not qualify as a limited quantum speedup due to the problem of suboptimal annealing times.

This problem was finally overcome in Ref.~\cite{Albash:2017aa}, which for the first time demonstrates an optimal annealing time, and can thus make positive claims about limited quantum speedup. Previous studies could not find an optimal annealing time since a class of problem instances had not been identified for which the shortest available annealing time ($20\mu s$ in the DW2, $5\mu s$ in all other D-Wave devices) was sufficiently short to observe on optimum given the largest problem size that could be tested. Using the D-Wave 2000Q (DW2KQ) device (with $2027$ qubits) Ref.~\cite{Albash:2017aa} demonstrated a simple one-unit cell gadget Hamiltonian
which, when added randomly to a constant fraction of the unit cells on top of similar frustrated loop problems as in Ref.~\cite{DW2000Q}, resulted in the observation of an optimal annealing time for frustrated loops defined on the hardware graph (also when using the DW2X device), as well as for frustrated loops defined on the logical graph of unit cells (each unit cell then being bound together tightly as a pseudo-spin in the physical problem, modulo the gadget Hamiltonian). For the latter, logical-planted instances, the DW2KQ exhibited a statistically significant scaling advantage over both single-spin-flip SA and SVMC. These results amount to the first observation of a  limited quantum speedup, since the existence of an optimal annealing time was certified. However, this did not amount to an unqualified quantum speedup since the DW2KQ's scaling was worse than the HFS algorithm, unit-cell cluster-flip SA, and SQA, which was found to have the best scaling. Nevertheless, this result paves the way towards future demonstrations of problems with optimal annealing times and hence certifiable scaling, a necessary requirement for any type of scaling speedup claim. However, even this may not be sufficient since other quantities remain that must eventually be optimized, such as the annealing schedule, which is known to play a crucial role in provable quantum speedups (specifically the Grover search problem \cite{Roland:2002ul,RPL:10}), and can conversely be used to potentially overturn (limited) quantum speedup claims.

\subsection{Lessons}
What lessons may be gleaned from this story for future benchmarks of quantum annealing devices with limited or no error correction and hundreds or thousands of qubits? 
\begin{enumerate}
	\item It is vitally important to carefully account for resource use, lest one be led astray with a fake speedup. In particular, quantum annealing requires a demonstration of an optimal annealing time before any definitive conclusion can be drawn about a quantum speedup. More generally, optimizing all known free parameters is almost certainly necessary to demonstrate a quantum speedup which will hold up to scrutiny. 
	\item One must distinguish between different types of quantum speedup. Comparisons between a quantum computational device and a single other solver are inherently limited to a demonstration of a ``potential quantum speedup". To go further, one must be sure to compare performance against a suite of algorithms, in particular those that mimic the device to some degree (such as SA or SQA). A speedup against such solvers would be considered a ``limited quantum speedup". If there is a consensus about the solvers that are the best at the original task, then a speedup against such solvers would be considered an unqualified ``quantum speedup". This would be a game-changing result. 
	\item Users of such quantum computational devices should perform something akin to gauge averaging in order to effectively estimate performance, by averaging over many different mappings from the logical problem to physical states, at least so long as the devices are not fully error corrected. Given that there is no good distribution for problem hardness as a function of this ensemble of mappings, nonparameteric techniques are appropriate.\footnote{We suggest using a variant of the bootstrap, the Bayesian bootstrap, first introduced in Ref.~\cite{rubin1981bayesian}, which can be shown to be the limit in the case of negligible prior information or large amounts of data of a Dirichlet process. Thus, it is arguably the only bootstrap procedure which is well-founded on Bayesian grounds. It involves reweighting the observed data, much like every bootstrap, but rather than sampling from a Multinomial$(N,[1/N,\ldots,1/N])$ distribution as in the frequentist bootstrap, one instead samples from the related Dirichlet$(1,\ldots,1)$ distribution. The primary advantage of the Bayesian bootstrap is that, unlike the frequentist bootstrap, the weight assigned any element in the dataset is always positive, i.e., there are no reweighted data vectors which leave out a data element, whereas the frequentist bootstrap has probability $1/e$ of dropping any given data element in a reweighted sample.}
	\item Choice of benchmark problem is key, and should be made with an eye toward the day when classical machines are vastly outpaced by quantum devices. For example, the transition from random Ising problems to frustrated loop/planted solutions problems was forced by the need to have reasonable benchmarks for devices so large that classical systems cannot solve them in a human lifetime.
\end{enumerate}

So far we have only addressed the question of benchmarking without any attempt at error correction. Since it is impossible to achieve a scalable quantum speedup without some means of correcting for errors, while benchmarking native problems may give some indication of the abilities of quantum annealing by looking at relatively small problem sizes, work on error correction lays the foundation for potential lasting quantum advantages over classical computing.

\begin{figure}[t]
\begin{center}
\subfigure[]{\includegraphics[width=0.12\textwidth]{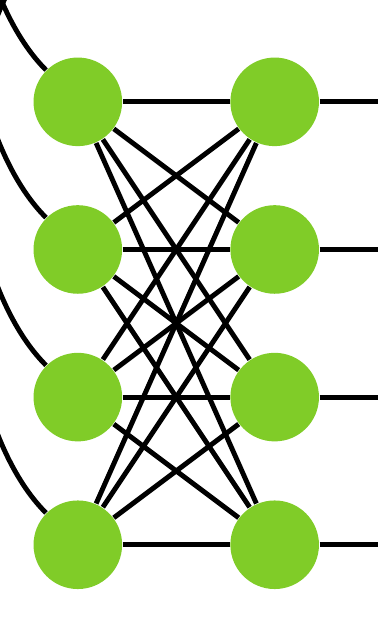}\label{fig:unitcell}} 
\subfigure[]{\includegraphics[width=0.12\textwidth]{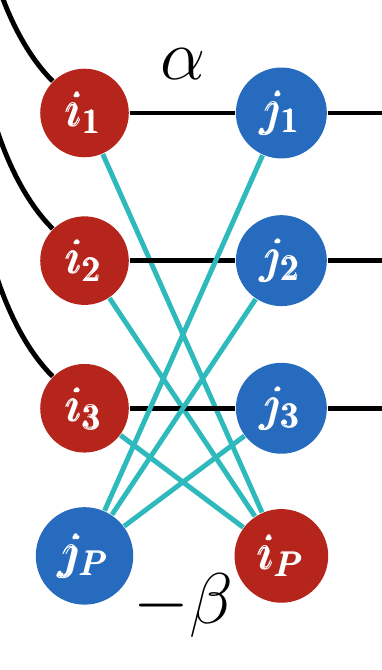}\label{fig:QACencoding}} 
\subfigure[]{\includegraphics[width=0.48\textwidth]{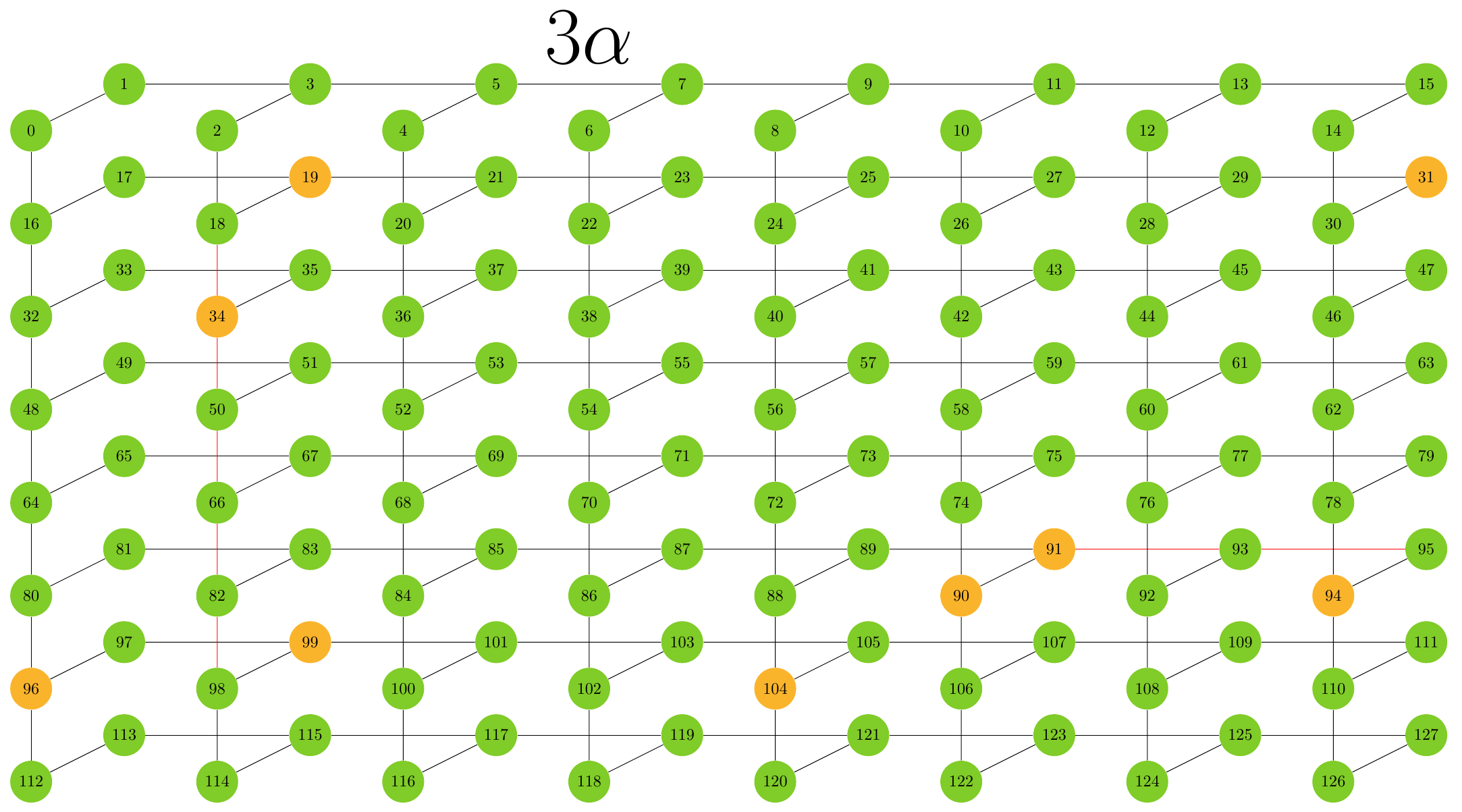}\label{fig:QACencodedHamiltonian}} 
\end{center}
\caption{The QAC unit cell and encoded graph introduced in Refs.~\cite{PAL:13,PAL:14}. (a) Schematic of one of the $64$ {unit cells} of the DW2 processor. Unit cells are arranged in an $8\times 8$ array forming a ``Chimera" graph between qubits. Each circle represents a physical qubit, and each line a programmable Ising coupling $\sigma^z_i\sigma^z_j$. Lines on the right (left) couple to the corresponding qubits in the neighboring unit cell to the right (above). 
(b) Two ``logical qubits" ($i$, red and $j$, blue) embedded within a single unit cell. Qubits labeled 1-3 are the ``problem qubits", the opposing qubit of the same color labeled $P$ is the ``penalty qubit". Problem qubits couple via the black lines with tunable strength $\alpha$ both inter- and intra-unit cell. Light blue lines of magnitude $\beta$ are ferromagnetic couplings between the problem qubits and their penalty qubit. (c) Encoded processor graph obtained from the Chimera graph by replacing each logical qubit by a circle. This is a non-planar graph with couplings of strength $3\alpha$. Green circles represent  complete logical qubits. Orange circles represent logical qubits lacking their penalty qubit. Red lines are {groups of} couplers that cannot {all} be simultaneously activated. 
}
\label{fig:QAC}
\end{figure}

%%%%%%
\begin{figure*}[ht]
\begin{center}
\subfigure[\ 2LG logical graph connectivity]{\includegraphics[width=0.32\textwidth]{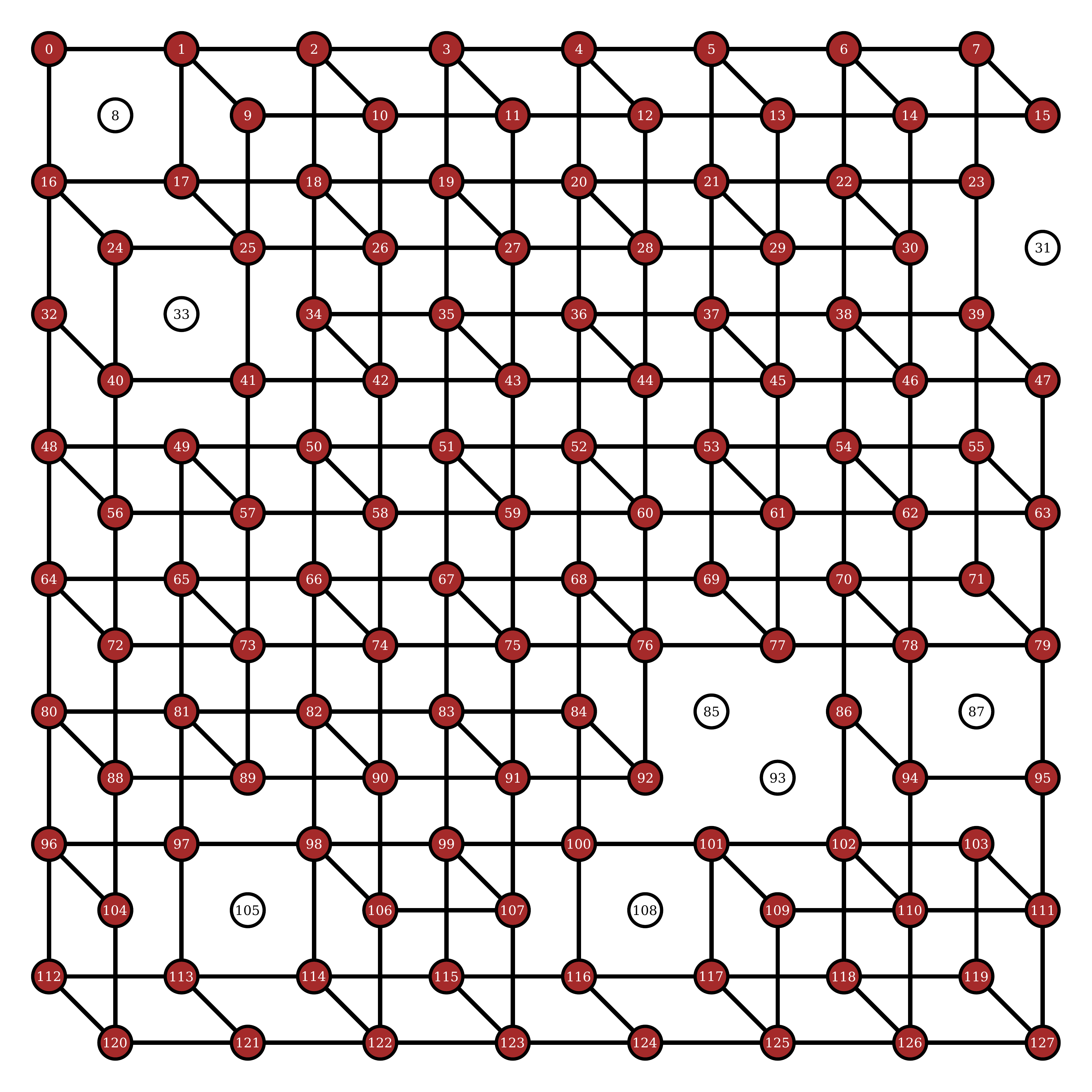}\label{fig:square-code-a}}
\subfigure[\ Minor embedding into Chimera graph]{\includegraphics[width=0.32\textwidth]{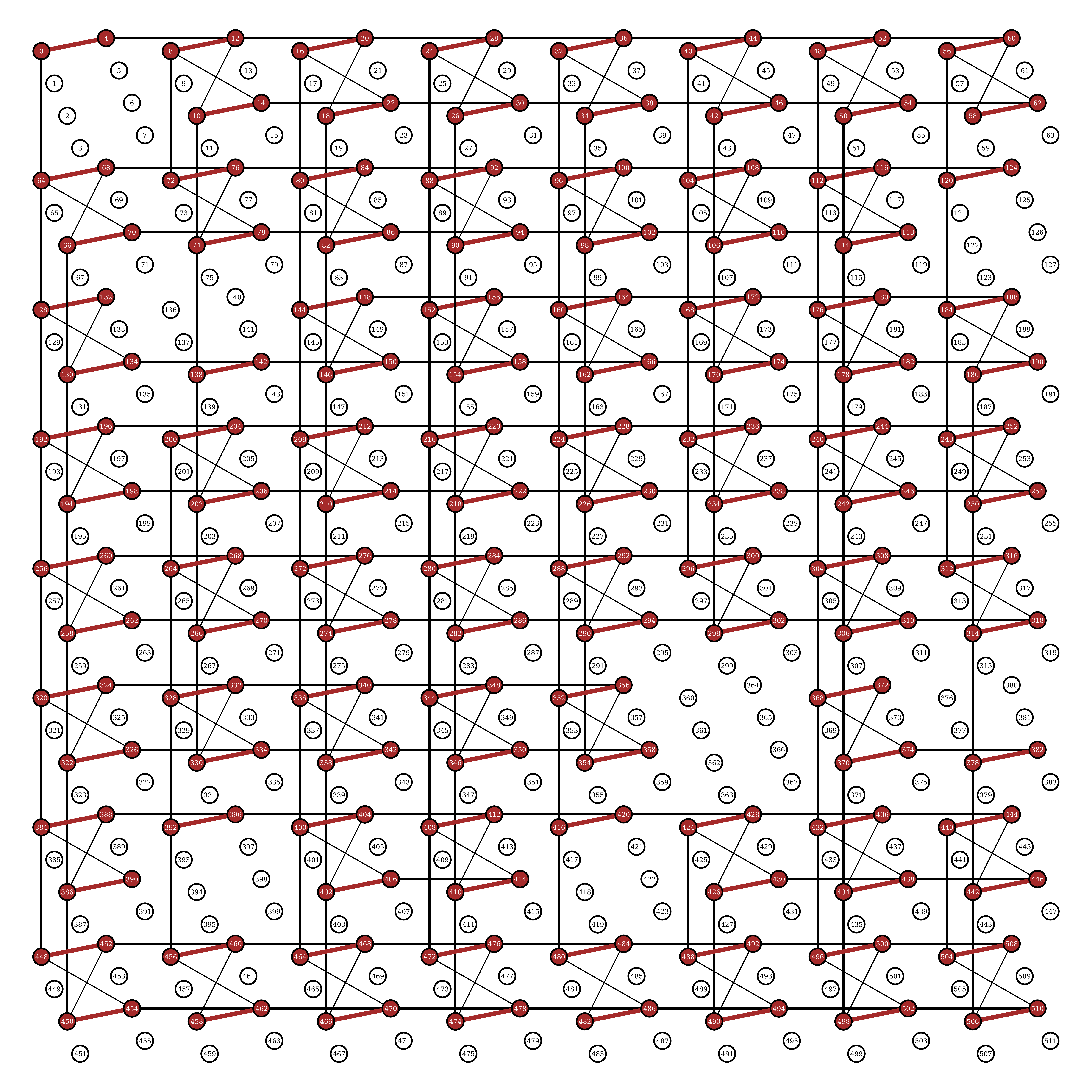}\label{fig:square-code-b}}
\subfigure[\ QAC-ME]{\includegraphics[width=0.32\textwidth]{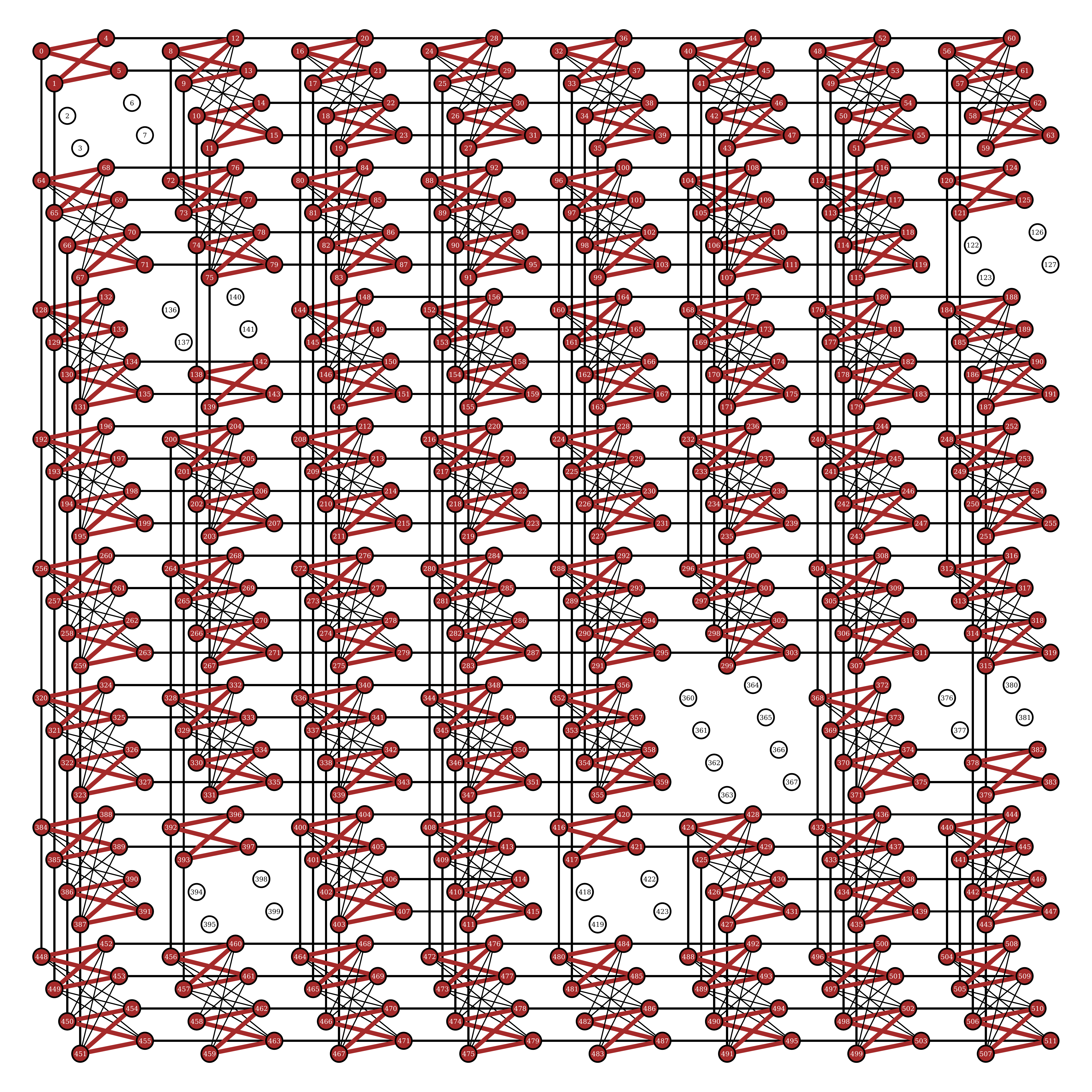}\label{fig:square-code-c}}
\caption{Illustration of mapping from a logical problem to a QAC-ME embedding, introduced in Ref.~\cite{Vinci:2015jt}. (a) The two-level-grid (2LG) graph, for which the Ising spin glass problem with couplings in $\{-1,0,1\}$ is NP-hard \cite{Barahona1982}. Disconnected vertices indicate spins with couplings set to zero, as well as  unused logical qubits in the logical DW2 Chimera graph. (b) Minor embedding of the 2LG graph into the physical DW2 Chimera graph. White circles correspond to unused or unusable qubits. (c) QAC-ME embedding of the 2LG problem using the ``square'' code.
In (b) and (c) penalties are represented by red (thick) couplings between groups of two (ME) and four (QAC-ME) physical qubits. Black (thin) links implement logical couplings.} 
\label{fig:square-code}
\end{center}
\end{figure*}
%%%%%%

\begin{figure}[t]
\begin{center}
\subfigure[\ Logical graph: 1st level.]{\includegraphics[width=0.23\textwidth]{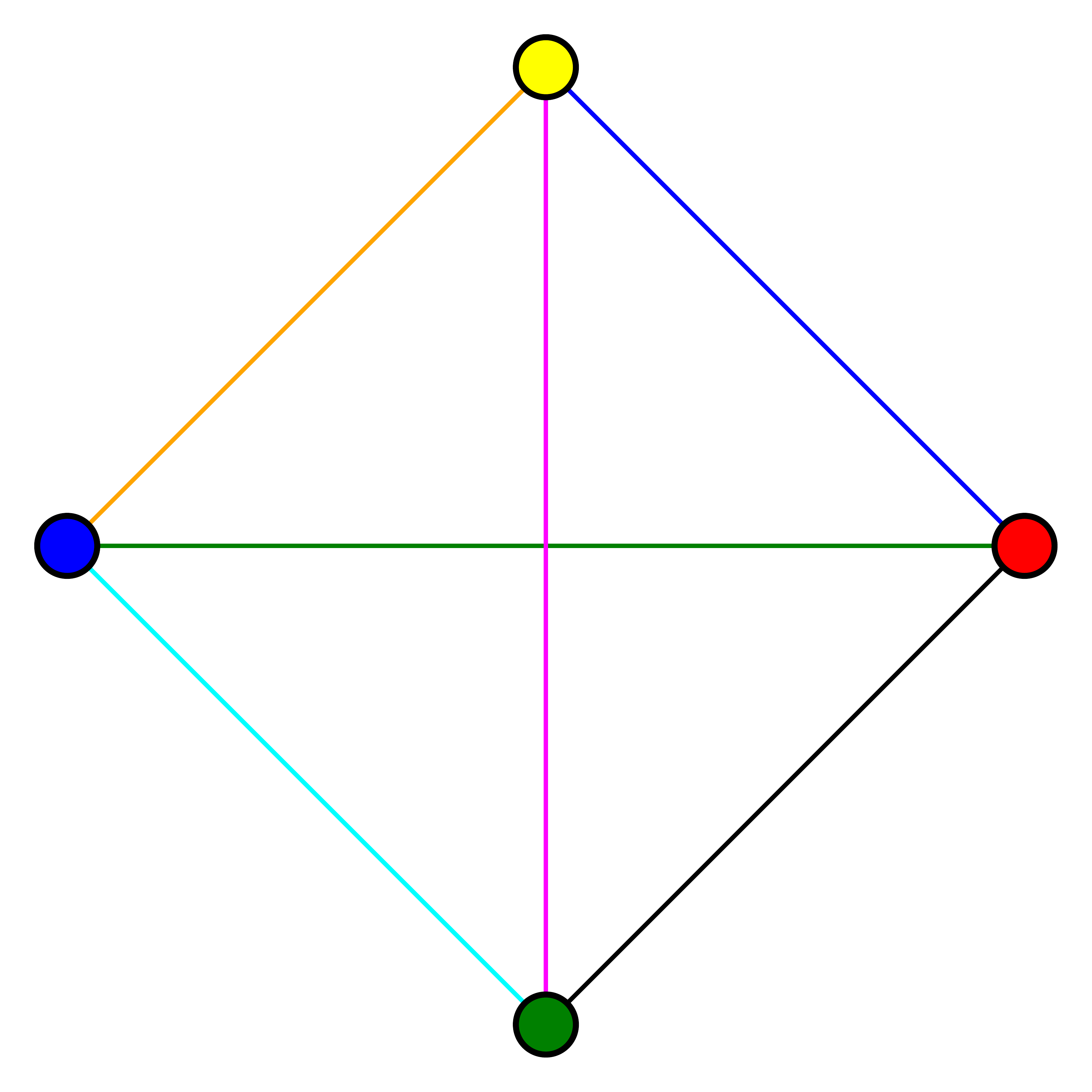}\label{fig:4x4_ideal_logical_plot_1}}
\subfigure[\ 1st level ME.]{\includegraphics[width=0.23\textwidth]{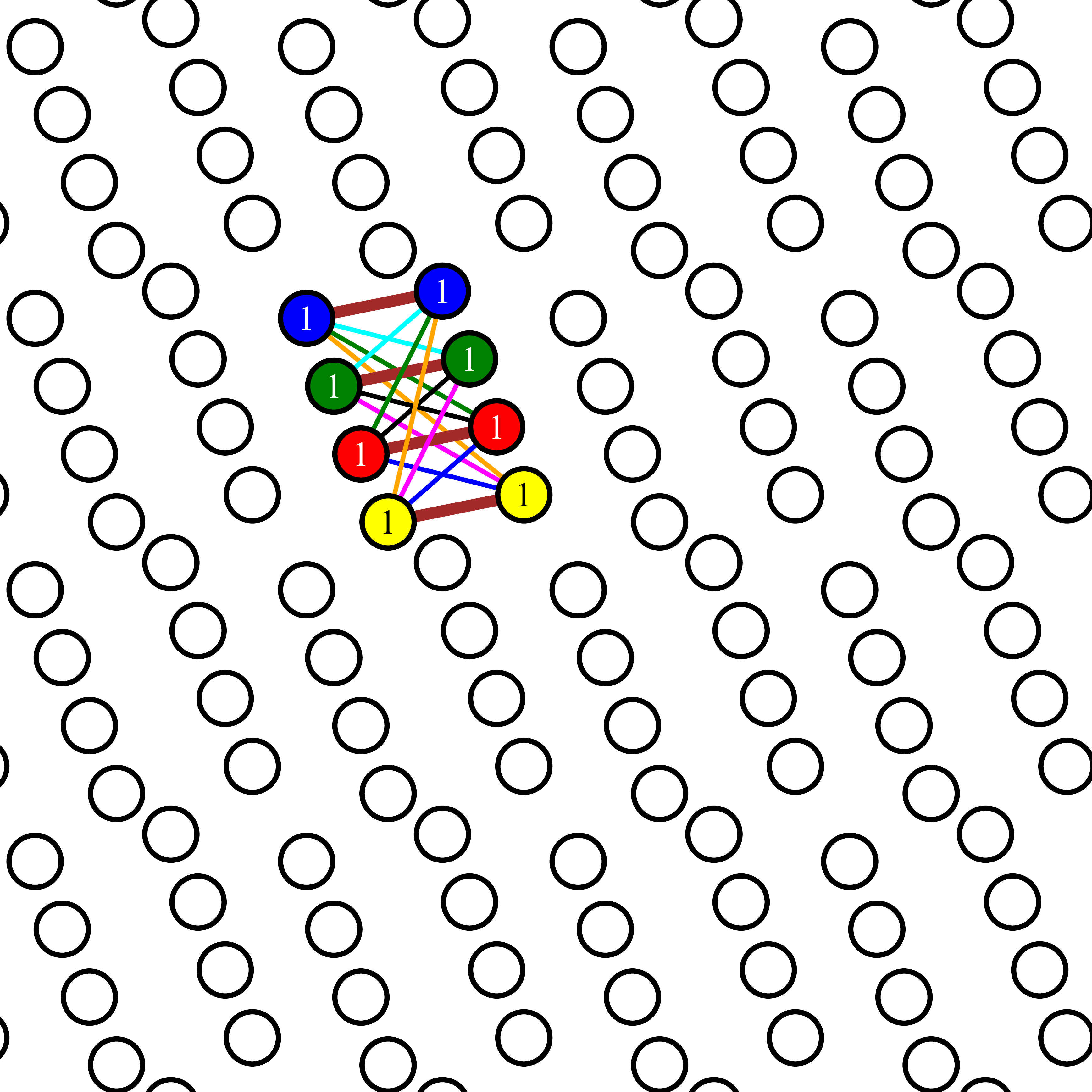}\label{fig:4x4_ideal_physical_1}}
\subfigure[\ Nested graph: 4th level.]{\includegraphics[width=0.23\textwidth]{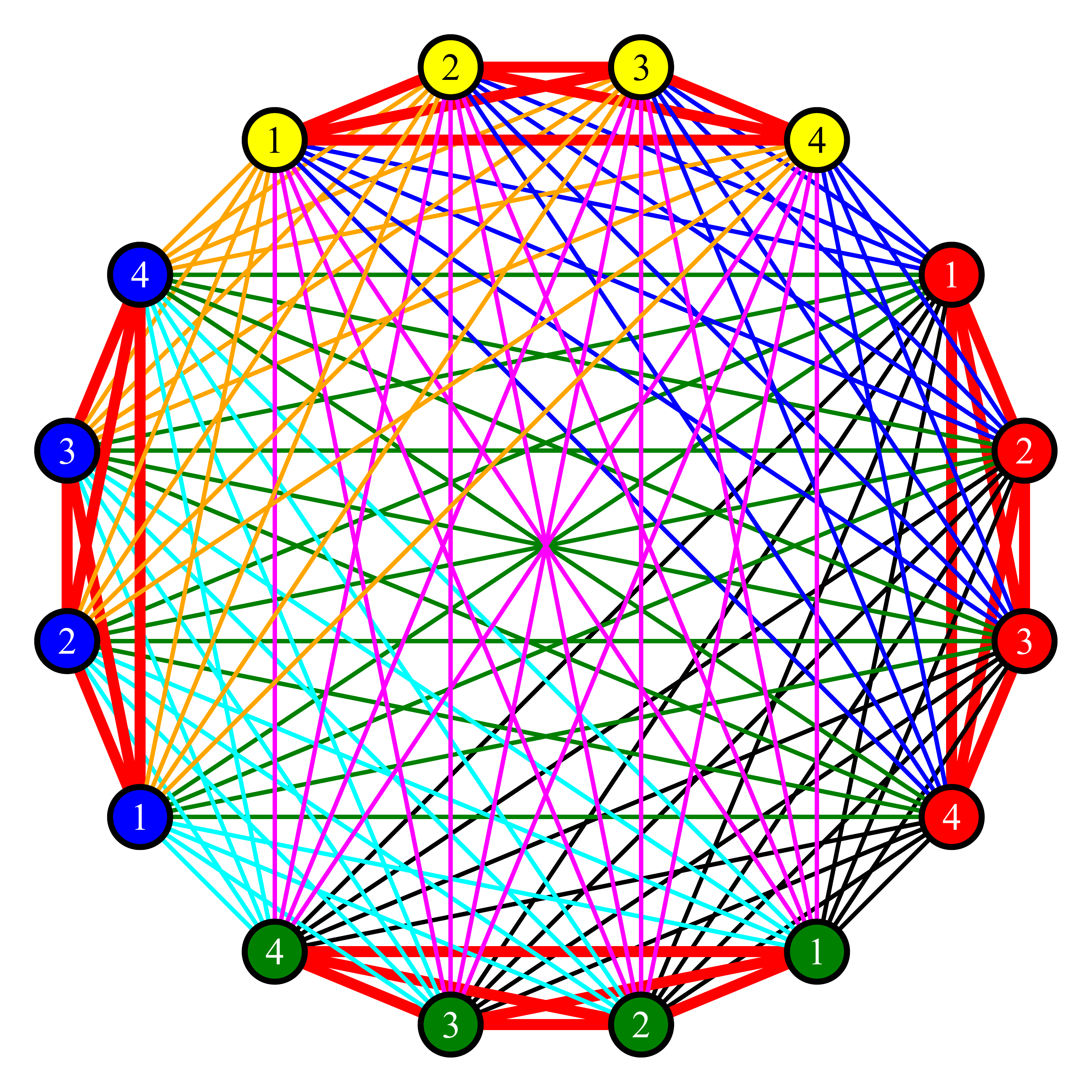}\label{fig:4x4_ideal_logical_plot_4}}
\subfigure[\ 4th level ME.]{\includegraphics[width=0.23\textwidth]{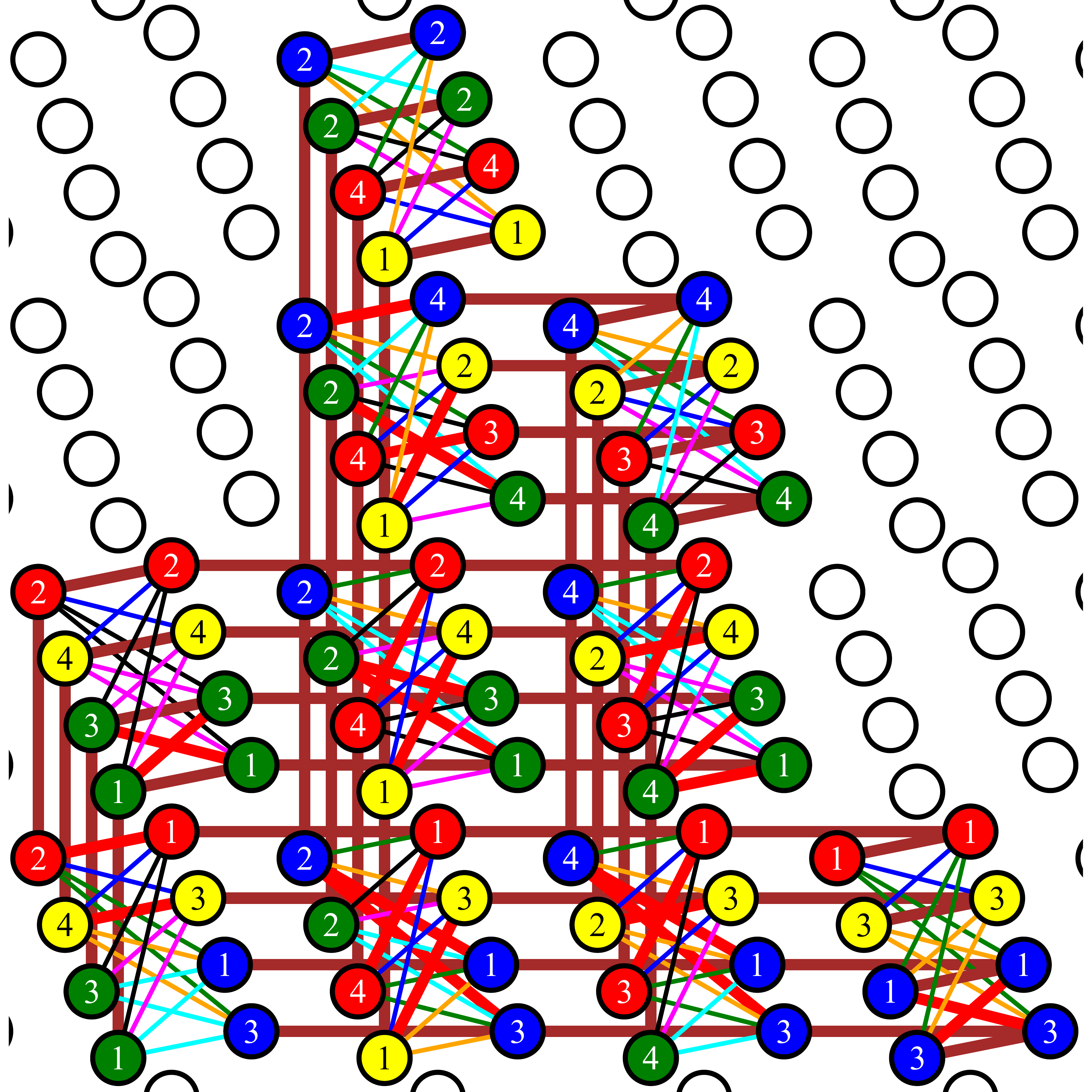}\label{fig:4x4_ideal_physical_4}}
\caption{Illustration of the nested QAC scheme introduced in Ref.~\cite{vinci2015nested}. In the left column, a  $C$-level nested graph is constructed by embedding a $K_N$ into a $K_{C\times N}$, with $N=4$ and $C=1$ (top) and $C=4$ (bottom). Red, thick 
couplers are energy penalties defined on the nested graph between the $(i,c)$ nested copies of each logical qubit $i$. 
The right column shows the nested graphs after minor embedding (ME) on the DW2X Chimera graph. Brown, thick couplers correspond to the ferromagnetic chains introduced in the process.
} 
\label{fig:NQAC}
\end{center}
\end{figure}
%%%%%%

\section{Error Correction}
\label{sec:err-corr}
The discussion of error correction in gate-based quantum computing is usually dominated by questions of fault tolerance and the error thresholds on one and two-qubit gates necessary to satisfy the fault tolerance theorems \cite{Aliferis:05,Raussendorf28092012,Lidar-Brun:book}. The state for quantum annealing and adiabatic quantum computing is quite different, as there is currently no known mechanism for achieving fault tolerance in such devices. Without the benefit of fault tolerance theorems, the best techniques we have available for managing errors in AQC and QA are energy gap protection \cite{jordan2006error}, dynamical decoupling \cite{PhysRevLett.100.160506}, and the Zeno effect \cite{PhysRevLett.108.080501}, three intimately related techniques \cite{Young:13}. Work on error correction in physically realized quantum annealers has focused on energy gap protection, as techniques for dynamical decoupling are unavailable in current generations of annealers due to the associated high bandwidth requirements. Both of these techniques are really more error suppression than correction, as rather then correct errors  they can lower the probability that errors will occur and mitigate their consequences should they happen.

The first demonstration of error suppression via energy energy gap protection in quantum annealers came with the introduction of the technique of quantum annealing correction (QAC) \cite{PAL:13}, which as the name suggests, also includes an active error correction component. In addition, Ref.~\cite{PAL:13} introduced a method for energy boosting by encoding the final Hamiltonian of a quantum annealing algorithm via multiple copies of the logical Hamiltonian operating on separate sets of physical qubits. This is in effect a simple classical repetition code. The copies are bound together by a penalty qubit whose action is to increase the energy of states of the physical system in which the copies are not in alignment with each other. The energy penalty for disagreement between the states effectively suppresses excitation to error states. Figure~\ref{fig:QAC} shows the structure of the encoding and the nature of the encoded problem graph on a DW2 device. The logical Hamiltonian is boosted to have an effective strength three times that achievable in the hardware by directly programming the Hamiltonian. A restriction of this approach is that only the final Hamiltonian is encoded, not the driver Hamiltonian $-\sum \sigma_i^x$, which means that while the final Hamiltonian's gap is significantly larger than in the unencoded case, it is difficult to verify that the minimum ground to first excited state gap of the quantum Hamiltonian is also enhanced. However, a mean field analysis shows that QAC softens the gap closure dependence on system size $N$, in the sense that for models exhibiting a first order quantum phase transition with the gap $\Delta$ scaling as $C^N$, the coefficient $0<C\leq 1$ grows monotonically with the QAC penalty strength $\gamma$, and saturates at $C=1$ for sufficiently large $\gamma$ \cite{MNAL:15,matsuura2017qac}. The same work also showed that after QAC, the free energy barrier between the global minimum and the local minimum the system is initially trapped in is reduced in both width and height for a variety of transverse field Ising spin models, including models with disorder such as the Hopfield model.

The encoding restriction on QAC is not intrinsic to the technique, but rather is the result of the lack of any higher order (more than single-body) $\sigma^x$ terms in the system Hamiltonian, which renders it impossible to form an effective logical $\sigma^x$ term. Decoding a QAC encoded Hamiltonian is as simple as applying a majority vote over the problem qubits. The method was tested in Ref.~\cite{PAL:13} on antiferromagnetic chains of various lengths, demonstrating a significant improvement in success probability of finding the ground state of the chains compared not only to the unencoded case but also the case of a four copy repetition code (since QAC uses four times the hardware resources, one could simply run four copies of the problem at once and pick the lowest energy solution from any copy). As implemented in Ref.~\cite{PAL:13}, the technique is not scalable (in the sense that both the energy boost and the gap against errors is constant), but it provided the first hope for systematically overcoming errors in the experimental quantum annealing. Another innovation was the use of many embeddings of the same logical Hamiltonian into compatible subgraphs, an idea which has found its way into the benchmarking context.

Chains have a trivial (classical) ground state, so a natural next test of QAC was to apply the method to random Ising problem instances \cite{PAL:14}. This provided a  demonstration that QAC could improve performance also on NP-hard problems defined on the QAC logical graph. Moreover, not only was the absolute performance on random Ising problems improved over both the unencoded and the classical repetition cases, but the scaling of the time to solution for those problems improved under QAC, with the caveat that no optimal annealing time was identified. In addition, QAC improved the robustness of the annealer to problem misspecification and increased the effective accuracy of the implemented problem Hamiltonian, as was shown by systematically reducing the energy scale of the final Hamiltonian. Since the hardware graph of the DW2 had missing qubits, encoding Ising problems required that some logical qubits went without a penalty qubit. Additional robustness was thus demonstrated by artificially increasing the number of penalty qubits lost: with up to $60\%$ of the logical qubits going without penalty qubits, QAC continued to work with negligible performance loss.

The next step in testing QAC was to apply it to minor-embedded problems \cite{Vinci:2015jt}, dubbed QAC-ME, thus going beyond natively embeddable problems such as chains and random Ising instances; see Fig.~\ref{fig:square-code}. A key innovation introduced in Ref.~\cite{Vinci:2015jt} is the introduction of non-uniform weights for both the QAC penalty terms as well as the strength of the chain in the minor embedding, making them both proportional to the mean coupling strength in their respective logical Hamiltonians. This was in part informed by previous minor embedding experiments such as Ref.~\cite{Venturelli:2014nx}, in which the optimal strength of the chain was found to be related to the emergence of the spin glass phase of the Hamiltonian. Since there is only a single strength $\sigma^x$ term applied to every qubit while the strength of the $\sigma^z$ terms depends on the choice of $h_i$ and $J_{ij}$, one can easily find that with a uniform penalty strength some qubits will ``freeze'' (i.e., no longer be effectively flipped by the driver Hamiltonian) before others, which can negatively impact solution quality. By locally fitting penalties to the strength of the logical problem Hamiltonian for each qubit, this process can be mitigated. Addressing decoding, Ref.~\cite{Vinci:2015jt} also proposed to use energy minimization, which involves directly minimizing the state of broken logical qubits (logical qubits whose physical qubits are not in alignment) given their neighboring qubits, and demonstrated that this can be done efficiently so long as the per-qubit probability of error is below the percolation threshold of the problem graph. And, going beyond the original QAC code of Ref.~\cite{PAL:13}, a new, scalable QAC ``square code" whose logical graph forms a two-level-grid was proposed in Ref.~\cite{Vinci:2015jt} (see Fig.~\ref{fig:square-code}). The square code has the attractive feature that it can be concatenated. To benchmark QAC-ME, the same kind of frustrated loop problems with planted solutions that were first introduced in Ref.~\cite{Hen:2015rt} were used. The results demonstrated a significant improvement in performance for non-uniform penalties over uniform penalty strengths, and that energy minimization was strongly preferable for decoding QAC-ME compared to standard majority vote decoding. The square code was compared with the original QAC code on chains in Ref.~\cite{Mishra:2015}.

Both the original QAC scheme and QAC-ME induce a graph of lower degree than that of the initial Hamiltonian. To overcome this and deal from the start with arbitrary Ising model Hamiltonians, a ``nested QAC" (NQAC) method was introduced in Ref.~\cite{vinci2015nested}. NQAC starts from a fully connected $K_N$ graph for the underlying problem and then maps this $N$ qubit problem into $C$ coupled copies of itself in a larger $K_{C\times N}$ graph. When run on a hardware graph of lower degree, this larger $K_{C\times N}$ graph is then minor-embedded, with the coupled copies doing the work of suppressing errors and helping to limit the formation of domain walls in the minor embedding chains. In this way, the number of physical qubits required to implement level-$C$ NQAC is approximately $C^2 N^2/4$. An illustration of the embedding of the $K_N$ Hamiltonian into the larger $K_{C\times N}$ Hamiltonian as well as a sample of the minor embedded graph are given in Fig.~\ref{fig:NQAC}.

The key finding of Ref.~\cite{vinci2015nested} is that NQAC effectively rescales the temperature of the system down by a factor of $\mu_C$ for $C$ nesting levels, with the theoretical expectation for a fully thermalized state being that $\mu_C\propto C^2$, based on a mean-field analysis. In practice, the scaling is not quite that fast, instead $\mu_C\approx C^{1.4}$ once the energy penalty tying the $C$ copies of the problem Hamiltonian together and the chain strength are optimized. This result is important since it means that one can trade qubits for an effective temperature reduction, that is controllable via the nesting level $C$. This suggests, at least in principle, that the effective temperature can be kept below the gap. For the DW2 device, for $C\geq 3$ NQAC was no longer able to improve the success probability more than classical repetition of the $C=1$ case, though this is probably because the base problem (a random antiferromagnetic Ising $K_8$ with $J_{ij}\in\{0.1,0.2,\dots,1\}$) was too easy. Tests on more advanced processors such as the DW2KQ and beyond will reveal whether techniques like NQAC hold at least part of the key to scalable quantum error suppression and correction on quantum annealing platforms, even if it and similar pure error suppression techniques will never be able to achieve true fault tolerance.

The story of error suppression and correction in experimental quantum annealing algorithms is one of a sequence of developments, building off both the results and lessons of native benchmarking work and the initial insight of energy gap protection as a fruitful and feasible path on current systems for error suppression, and generalizing to more and more useful encoded graphs, until finally reaching NQAC with its fully general encoded Ising Hamiltonian and potential for arbitrarily large and strong error suppression. Future paths for investigation of experimental quantum annealing correction center on expanded benchmarking of NQAC for larger systems and for application-domain problems, as well as continued theoretical development of error suppression and correction techniques. For example, using subsystem codes it is possible to construct error suppression schemes appropriate for adiabatic quantum computing that use only two-body interactions, so that by adding $\sigma^x_i \sigma^x_j$ terms to the driver Hamiltonian one could significantly improve over the current state of the art of QAC  \cite{Marvian-Lidar:16,Jiang:2015kx}.

\section{Conclusion}
As the field of quantum computing, and quantum annealing in particular, expands rapidly and the number of available platforms rises, methods to validate the fidelity of the platform to its stated physical model, verifying entanglement and tunneling, strong benchmarking methods, and error correction/suppression techniques will be vital in discerning which platforms truly can offer advantages over classical computation. Several years have been spent developing methods to meet each of these challenges, particularly targeting existing quantum annealers, but many of these methods can be readily adapted to other systems that implement programmable Ising model Hamiltonians \cite{Inagaki:2016aa}. Insights from each of these areas are informing development in the others. For example, insight into and experience with small gadgets from quantum validation informed recent work demonstrating a limited quantum speedup on D-Wave quantum annealers using a small gadget to generate an optimal annealing time on existing machines, while insights from benchmarking regularly inform developments in error suppression and correction. Work on benchmarking has provided guidelines and methods of analysis which can be used by anyone seeking to characterize the performance of a putative quantum computing device, while error suppression work has laid the foundation for more extensive experiments and solving more difficult problems on future, larger quantum annealing devices. Thus, one answer to the question of what one may want to use a $1000$-qubit quantum computer for, is in our view the type of bootstrapping we have reviewed in this article, where a productive interplay among quantum validation testing, benchmarking, and error correction has led to a sequence of advances that will inform even larger quantum computation experiments, until one day a test drive with a brand new quantum computer will take us to the ultimate destination of undisputed quantum supremacy and unqualifed quantum speedup.

\vspace{.8cm}

\acknowledgements
We are grateful to the many colleagues with whom we have collaborated over the past several years of test-driving quantum annealers, and in particular to Tameem Albash for helpful discussions and comments on this article. This work was supported under ARO grant number W911NF-12-1-0523, ARO MURI grant No. W911NF-15-1-0582, and NSF grant number INSPIRE-1551064.


\begin{thebibliography}{101}%
\makeatletter
\providecommand \@ifxundefined [1]{%
 \@ifx{#1\undefined}
}%
\providecommand \@ifnum [1]{%
 \ifnum #1\expandafter \@firstoftwo
 \else \expandafter \@secondoftwo
 \fi
}%
\providecommand \@ifx [1]{%
 \ifx #1\expandafter \@firstoftwo
 \else \expandafter \@secondoftwo
 \fi
}%
\providecommand \natexlab [1]{#1}%
\providecommand \enquote  [1]{``#1''}%
\providecommand \bibnamefont  [1]{#1}%
\providecommand \bibfnamefont [1]{#1}%
\providecommand \citenamefont [1]{#1}%
\providecommand \href@noop [0]{\@secondoftwo}%
\providecommand \href [0]{\begingroup \@sanitize@url \@href}%
\providecommand \@href[1]{\@@startlink{#1}\@@href}%
\providecommand \@@href[1]{\endgroup#1\@@endlink}%
\providecommand \@sanitize@url [0]{\catcode `\\12\catcode `\$12\catcode
  `\&12\catcode `\#12\catcode `\^12\catcode `\_12\catcode `\%12\relax}%
\providecommand \@@startlink[1]{}%
\providecommand \@@endlink[0]{}%
\providecommand \url  [0]{\begingroup\@sanitize@url \@url }%
\providecommand \@url [1]{\endgroup\@href {#1}{\urlprefix }}%
\providecommand \urlprefix  [0]{URL }%
\providecommand \Eprint [0]{\href }%
\providecommand \doibase [0]{http://dx.doi.org/}%
\providecommand \selectlanguage [0]{\@gobble}%
\providecommand \bibinfo  [0]{\@secondoftwo}%
\providecommand \bibfield  [0]{\@secondoftwo}%
\providecommand \translation [1]{[#1]}%
\providecommand \BibitemOpen [0]{}%
\providecommand \bibitemStop [0]{}%
\providecommand \bibitemNoStop [0]{.\EOS\space}%
\providecommand \EOS [0]{\spacefactor3000\relax}%
\providecommand \BibitemShut  [1]{\csname bibitem#1\endcsname}%
\let\auto@bib@innerbib\@empty
%</preamble>
\bibitem [{\citenamefont {Reichardt}\ \emph {et~al.}(2013)\citenamefont
  {Reichardt}, \citenamefont {Unger},\ and\ \citenamefont
  {Vazirani}}]{Reichardt:2013db}%
  \BibitemOpen
  \bibfield  {author} {\bibinfo {author} {\bibfnamefont {B.~W.}\ \bibnamefont
  {Reichardt}}, \bibinfo {author} {\bibfnamefont {F.}~\bibnamefont {Unger}}, \
  and\ \bibinfo {author} {\bibfnamefont {U.}~\bibnamefont {Vazirani}},\ }\href
  {\doibase 10.1038/nature12035} {\bibfield  {journal} {\bibinfo  {journal}
  {Nature}\ }\textbf {\bibinfo {volume} {496}},\ \bibinfo {pages} {456}
  (\bibinfo {year} {2013})}\BibitemShut {NoStop}%
\bibitem [{\citenamefont {Chuang}\ and\ \citenamefont
  {Nielsen}(1997)}]{Chuang:97c}%
  \BibitemOpen
  \bibfield  {author} {\bibinfo {author} {\bibfnamefont {I.~L.}\ \bibnamefont
  {Chuang}}\ and\ \bibinfo {author} {\bibfnamefont {M.~A.}\ \bibnamefont
  {Nielsen}},\ }\bibfield  {booktitle} {\emph {\bibinfo {booktitle} {Journal of
  Modern Optics}},\ }\href
  {http://www.tandfonline.com/doi/abs/10.1080/09500349708231894} {\bibfield
  {journal} {\bibinfo  {journal} {Journal of Modern Optics}\ }\textbf {\bibinfo
  {volume} {44}},\ \bibinfo {pages} {2455} (\bibinfo {year}
  {1997})}\BibitemShut {NoStop}%
\bibitem [{\citenamefont {Mohseni}\ \emph {et~al.}(2008)\citenamefont
  {Mohseni}, \citenamefont {Rezakhani},\ and\ \citenamefont
  {Lidar}}]{Mohseni:2008ly}%
  \BibitemOpen
  \bibfield  {author} {\bibinfo {author} {\bibfnamefont {M.}~\bibnamefont
  {Mohseni}}, \bibinfo {author} {\bibfnamefont {A.~T.}\ \bibnamefont
  {Rezakhani}}, \ and\ \bibinfo {author} {\bibfnamefont {D.~A.}\ \bibnamefont
  {Lidar}},\ }\href {http://link.aps.org/doi/10.1103/PhysRevA.77.032322}
  {\bibfield  {journal} {\bibinfo  {journal} {Phys. Rev. A}\ }\textbf {\bibinfo
  {volume} {77}},\ \bibinfo {pages} {032322} (\bibinfo {year}
  {2008})}\BibitemShut {NoStop}%
\bibitem [{\citenamefont {Blume-Kohout}\ \emph {et~al.}(2013)\citenamefont
  {Blume-Kohout}, \citenamefont {Gamble}, \citenamefont {Nielsen},
  \citenamefont {Mizrahi}, \citenamefont {Sterk},\ and\ \citenamefont
  {Maunz}}]{blume2013robust}%
  \BibitemOpen
  \bibfield  {author} {\bibinfo {author} {\bibfnamefont {R.}~\bibnamefont
  {Blume-Kohout}}, \bibinfo {author} {\bibfnamefont {J.~K.}\ \bibnamefont
  {Gamble}}, \bibinfo {author} {\bibfnamefont {E.}~\bibnamefont {Nielsen}},
  \bibinfo {author} {\bibfnamefont {J.}~\bibnamefont {Mizrahi}}, \bibinfo
  {author} {\bibfnamefont {J.~D.}\ \bibnamefont {Sterk}}, \ and\ \bibinfo
  {author} {\bibfnamefont {P.}~\bibnamefont {Maunz}},\ }\href
  {http://arXiv.org/abs/1310.4492} {\bibfield  {journal} {\bibinfo  {journal}
  {arXiv:1310.4492}\ } (\bibinfo {year} {2013})}\BibitemShut {NoStop}%
\bibitem [{\citenamefont {Greenbaum}(2015)}]{Greenbaum:2015aa}%
  \BibitemOpen
  \bibfield  {author} {\bibinfo {author} {\bibfnamefont {D.}~\bibnamefont
  {Greenbaum}},\ }\href {http://arXiv.org/abs/1509.02921} {\bibfield  {journal}
  {\bibinfo  {journal} {arXiv:1509.02921}\ } (\bibinfo {year}
  {2015})}\BibitemShut {NoStop}%
\bibitem [{\citenamefont {Childs}\ \emph {et~al.}(2001)\citenamefont {Childs},
  \citenamefont {Chuang},\ and\ \citenamefont {Leung}}]{Childs:00}%
  \BibitemOpen
  \bibfield  {author} {\bibinfo {author} {\bibfnamefont {A.~M.}\ \bibnamefont
  {Childs}}, \bibinfo {author} {\bibfnamefont {I.~L.}\ \bibnamefont {Chuang}},
  \ and\ \bibinfo {author} {\bibfnamefont {D.~W.}\ \bibnamefont {Leung}},\
  }\href {https://link.aps.org/doi/10.1103/PhysRevA.64.012314} {\bibfield
  {journal} {\bibinfo  {journal} {Physical Review A}\ }\textbf {\bibinfo
  {volume} {64}},\ \bibinfo {pages} {012314} (\bibinfo {year}
  {2001})}\BibitemShut {NoStop}%
\bibitem [{\citenamefont {Blume-Kohout}\ \emph {et~al.}(2017)\citenamefont
  {Blume-Kohout}, \citenamefont {Gamble}, \citenamefont {Nielsen},
  \citenamefont {Rudinger}, \citenamefont {Mizrahi}, \citenamefont {Fortier},\
  and\ \citenamefont {Maunz}}]{Blume-Kohout:2017aa}%
  \BibitemOpen
  \bibfield  {author} {\bibinfo {author} {\bibfnamefont {R.}~\bibnamefont
  {Blume-Kohout}}, \bibinfo {author} {\bibfnamefont {J.~K.}\ \bibnamefont
  {Gamble}}, \bibinfo {author} {\bibfnamefont {E.}~\bibnamefont {Nielsen}},
  \bibinfo {author} {\bibfnamefont {K.}~\bibnamefont {Rudinger}}, \bibinfo
  {author} {\bibfnamefont {J.}~\bibnamefont {Mizrahi}}, \bibinfo {author}
  {\bibfnamefont {K.}~\bibnamefont {Fortier}}, \ and\ \bibinfo {author}
  {\bibfnamefont {P.}~\bibnamefont {Maunz}},\ }\href
  {http://dx.doi.org/10.1038/ncomms14485} {\bibfield  {journal} {\bibinfo
  {journal} {Nature Communications}\ }\textbf {\bibinfo {volume} {8}},\
  \bibinfo {pages} {14485} (\bibinfo {year} {2017})}\BibitemShut {NoStop}%
\bibitem [{\citenamefont {Kadowaki}\ and\ \citenamefont
  {Nishimori}(1998)}]{kadowaki_quantum_1998}%
  \BibitemOpen
  \bibfield  {author} {\bibinfo {author} {\bibfnamefont {T.}~\bibnamefont
  {Kadowaki}}\ and\ \bibinfo {author} {\bibfnamefont {H.}~\bibnamefont
  {Nishimori}},\ }\href
  {http://journals.aps.org/pre/abstract/10.1103/PhysRevE.58.5355} {\bibfield
  {journal} {\bibinfo  {journal} {Phys. Rev. E}\ }\textbf {\bibinfo {volume}
  {58}},\ \bibinfo {pages} {5355} (\bibinfo {year} {1998})}\BibitemShut
  {NoStop}%
\bibitem [{\citenamefont {Das}\ and\ \citenamefont
  {Chakrabarti}(2008)}]{RevModPhys.80.1061}%
  \BibitemOpen
  \bibfield  {author} {\bibinfo {author} {\bibfnamefont {A.}~\bibnamefont
  {Das}}\ and\ \bibinfo {author} {\bibfnamefont {B.~K.}\ \bibnamefont
  {Chakrabarti}},\ }\href {\doibase 10.1103/RevModPhys.80.1061} {\bibfield
  {journal} {\bibinfo  {journal} {Rev. Mod. Phys.}\ }\textbf {\bibinfo {volume}
  {80}},\ \bibinfo {pages} {1061} (\bibinfo {year} {2008})}\BibitemShut
  {NoStop}%
\bibitem [{\citenamefont {Farhi}\ \emph {et~al.}(2001)\citenamefont {Farhi},
  \citenamefont {Goldstone}, \citenamefont {Gutmann}, \citenamefont {Lapan},
  \citenamefont {Lundgren},\ and\ \citenamefont {Preda}}]{farhi2001quantum}%
  \BibitemOpen
  \bibfield  {author} {\bibinfo {author} {\bibfnamefont {E.}~\bibnamefont
  {Farhi}}, \bibinfo {author} {\bibfnamefont {J.}~\bibnamefont {Goldstone}},
  \bibinfo {author} {\bibfnamefont {S.}~\bibnamefont {Gutmann}}, \bibinfo
  {author} {\bibfnamefont {J.}~\bibnamefont {Lapan}}, \bibinfo {author}
  {\bibfnamefont {A.}~\bibnamefont {Lundgren}}, \ and\ \bibinfo {author}
  {\bibfnamefont {D.}~\bibnamefont {Preda}},\ }\href
  {http://science.sciencemag.org/content/292/5516/472} {\bibfield  {journal}
  {\bibinfo  {journal} {Science}\ }\textbf {\bibinfo {volume} {292}},\ \bibinfo
  {pages} {472} (\bibinfo {year} {2001})}\BibitemShut {NoStop}%
\bibitem [{\citenamefont {{Kaminsky}}\ and\ \citenamefont
  {{Lloyd}}(2004)}]{2002quant.ph.11152K}%
  \BibitemOpen
  \bibfield  {author} {\bibinfo {author} {\bibfnamefont {W.~M.}\ \bibnamefont
  {{Kaminsky}}}\ and\ \bibinfo {author} {\bibfnamefont {S.}~\bibnamefont
  {{Lloyd}}},\ }in\ \href {http://arxiv.org/abs/quant-ph/0211152} {\emph
  {\bibinfo {booktitle} {Quantum Computing and Quantum Bits in Mesoscopic
  Systems}}},\ \bibinfo {editor} {edited by\ \bibinfo {editor} {\bibfnamefont
  {A.}~\bibnamefont {Leggett}}, \bibinfo {editor} {\bibfnamefont
  {B.}~\bibnamefont {Ruggiero}}, \ and\ \bibinfo {editor} {\bibfnamefont
  {P.}~\bibnamefont {Silvestrini}}}\ (\bibinfo  {publisher} {Kluwer
  Academic/Plenum Publ.},\ \bibinfo {year} {2004})\ \Eprint
  {http://arxiv.org/abs/arXiv:quant-ph/0211152} {arXiv:quant-ph/0211152}
  \BibitemShut {NoStop}%
\bibitem [{\citenamefont {Kaminsky}\ \emph {et~al.}(2004)\citenamefont
  {Kaminsky}, \citenamefont {Lloyd},\ and\ \citenamefont
  {Orlando}}]{Kaminsky:2004fk}%
  \BibitemOpen
  \bibfield  {author} {\bibinfo {author} {\bibfnamefont {W.~M.}\ \bibnamefont
  {Kaminsky}}, \bibinfo {author} {\bibfnamefont {S.}~\bibnamefont {Lloyd}}, \
  and\ \bibinfo {author} {\bibfnamefont {T.~P.}\ \bibnamefont {Orlando}},\
  }\href {http://arXiv.org/abs/quant-ph/0403090} {\bibfield  {journal}
  {\bibinfo  {journal} {arXiv:quant-ph/0403090}\ } (\bibinfo {year}
  {2004})}\BibitemShut {NoStop}%
\bibitem [{\citenamefont {Albash}\ and\ \citenamefont
  {Lidar}(2016)}]{Albash-Lidar:RMP}%
  \BibitemOpen
  \bibfield  {author} {\bibinfo {author} {\bibfnamefont {T.}~\bibnamefont
  {Albash}}\ and\ \bibinfo {author} {\bibfnamefont {D.~A.}\ \bibnamefont
  {Lidar}},\ }\href {http://arXiv.org/abs/1611.04471} {\bibfield  {journal}
  {\bibinfo  {journal} {arXiv:1611.04471}\ } (\bibinfo {year}
  {2016})}\BibitemShut {NoStop}%
\bibitem [{\citenamefont {Berkley}\ \emph {et~al.}(2013)\citenamefont
  {Berkley}, \citenamefont {Przybysz}, \citenamefont {Lanting}, \citenamefont
  {Harris}, \citenamefont {Dickson}, \citenamefont {Altomare}, \citenamefont
  {Amin}, \citenamefont {Bunyk}, \citenamefont {Enderud}, \citenamefont
  {Hoskinson}, \citenamefont {Johnson}, \citenamefont {Ladizinsky},
  \citenamefont {Neufeld}, \citenamefont {Rich}, \citenamefont {Smirnov},
  \citenamefont {Tolkacheva}, \citenamefont {Uchaikin},\ and\ \citenamefont
  {Wilson}}]{Berkley:2013bf}%
  \BibitemOpen
  \bibfield  {author} {\bibinfo {author} {\bibfnamefont {A.~J.}\ \bibnamefont
  {Berkley}}, \bibinfo {author} {\bibfnamefont {A.~J.}\ \bibnamefont
  {Przybysz}}, \bibinfo {author} {\bibfnamefont {T.}~\bibnamefont {Lanting}},
  \bibinfo {author} {\bibfnamefont {R.}~\bibnamefont {Harris}}, \bibinfo
  {author} {\bibfnamefont {N.}~\bibnamefont {Dickson}}, \bibinfo {author}
  {\bibfnamefont {F.}~\bibnamefont {Altomare}}, \bibinfo {author}
  {\bibfnamefont {M.~H.}\ \bibnamefont {Amin}}, \bibinfo {author}
  {\bibfnamefont {P.}~\bibnamefont {Bunyk}}, \bibinfo {author} {\bibfnamefont
  {C.}~\bibnamefont {Enderud}}, \bibinfo {author} {\bibfnamefont
  {E.}~\bibnamefont {Hoskinson}}, \bibinfo {author} {\bibfnamefont {M.~W.}\
  \bibnamefont {Johnson}}, \bibinfo {author} {\bibfnamefont {E.}~\bibnamefont
  {Ladizinsky}}, \bibinfo {author} {\bibfnamefont {R.}~\bibnamefont {Neufeld}},
  \bibinfo {author} {\bibfnamefont {C.}~\bibnamefont {Rich}}, \bibinfo {author}
  {\bibfnamefont {A.~Y.}\ \bibnamefont {Smirnov}}, \bibinfo {author}
  {\bibfnamefont {E.}~\bibnamefont {Tolkacheva}}, \bibinfo {author}
  {\bibfnamefont {S.}~\bibnamefont {Uchaikin}}, \ and\ \bibinfo {author}
  {\bibfnamefont {A.~B.}\ \bibnamefont {Wilson}},\ }\href
  {http://link.aps.org/doi/10.1103/PhysRevB.87.020502} {\bibfield  {journal}
  {\bibinfo  {journal} {Phys. Rev. B}\ }\textbf {\bibinfo {volume} {87}},\
  \bibinfo {pages} {020502} (\bibinfo {year} {2013})}\BibitemShut {NoStop}%
\bibitem [{\citenamefont {Preskill}(2012)}]{Preskill:2012aa}%
  \BibitemOpen
  \bibfield  {author} {\bibinfo {author} {\bibfnamefont {J.}~\bibnamefont
  {Preskill}},\ }\href {http://arXiv.org/abs/1203.5813} {\bibfield  {journal}
  {\bibinfo  {journal} {arXiv:1203.5813}\ } (\bibinfo {year}
  {2012})}\BibitemShut {NoStop}%
\bibitem [{\citenamefont {Flammia}(2017)}]{Qsupremacy-debate}%
  \BibitemOpen
  \bibfield  {author} {\bibinfo {author} {\bibfnamefont {S.}~\bibnamefont
  {Flammia}},\ }\href {http://dabacon.org/pontiff/?p=11863} {\enquote {\bibinfo
  {title} {{The Quantum Pontiff blog: ``Quantum Advantage"}},}\ } (\bibinfo
  {year} {2017})\BibitemShut {NoStop}%
\bibitem [{\citenamefont {Aaronson}\ and\ \citenamefont
  {Chen}(2016)}]{Aaronson:2016aa}%
  \BibitemOpen
  \bibfield  {author} {\bibinfo {author} {\bibfnamefont {S.}~\bibnamefont
  {Aaronson}}\ and\ \bibinfo {author} {\bibfnamefont {L.}~\bibnamefont
  {Chen}},\ }\href {http://arXiv.org/abs/1612.05903} {\bibfield  {journal}
  {\bibinfo  {journal} {arXiv:1612.05903}\ } (\bibinfo {year}
  {2016})}\BibitemShut {NoStop}%
\bibitem [{\citenamefont {Bremner}\ \emph {et~al.}(2016)\citenamefont
  {Bremner}, \citenamefont {Montanaro},\ and\ \citenamefont
  {Shepherd}}]{Bremner:2016aa}%
  \BibitemOpen
  \bibfield  {author} {\bibinfo {author} {\bibfnamefont {M.~J.}\ \bibnamefont
  {Bremner}}, \bibinfo {author} {\bibfnamefont {A.}~\bibnamefont {Montanaro}},
  \ and\ \bibinfo {author} {\bibfnamefont {D.~J.}\ \bibnamefont {Shepherd}},\
  }\href {https://link.aps.org/doi/10.1103/PhysRevLett.117.080501} {\bibfield
  {journal} {\bibinfo  {journal} {Physical Review Letters}\ }\textbf {\bibinfo
  {volume} {117}},\ \bibinfo {pages} {080501} (\bibinfo {year}
  {2016})}\BibitemShut {NoStop}%
\bibitem [{\citenamefont {Farhi}\ and\ \citenamefont
  {Harrow}(2016)}]{FarhiHarrow-QAOA}%
  \BibitemOpen
  \bibfield  {author} {\bibinfo {author} {\bibfnamefont {E.}~\bibnamefont
  {Farhi}}\ and\ \bibinfo {author} {\bibfnamefont {A.~W.}\ \bibnamefont
  {Harrow}},\ }\href {https://arxiv.org/abs/1602.07674} {\bibfield  {journal}
  {\bibinfo  {journal} {arXiv:1602.07674}\ } (\bibinfo {year}
  {2016})}\BibitemShut {NoStop}%
\bibitem [{\citenamefont {Gao}\ \emph {et~al.}(2017)\citenamefont {Gao},
  \citenamefont {Wang},\ and\ \citenamefont {Duan}}]{Gao:2017aa}%
  \BibitemOpen
  \bibfield  {author} {\bibinfo {author} {\bibfnamefont {X.}~\bibnamefont
  {Gao}}, \bibinfo {author} {\bibfnamefont {S.-T.}\ \bibnamefont {Wang}}, \
  and\ \bibinfo {author} {\bibfnamefont {L.~M.}\ \bibnamefont {Duan}},\ }\href
  {https://link.aps.org/doi/10.1103/PhysRevLett.118.040502} {\bibfield
  {journal} {\bibinfo  {journal} {Physical Review Letters}\ }\textbf {\bibinfo
  {volume} {118}},\ \bibinfo {pages} {040502} (\bibinfo {year}
  {2017})}\BibitemShut {NoStop}%
\bibitem [{\citenamefont {Boixo}\ \emph
  {et~al.}(2016{\natexlab{a}})\citenamefont {Boixo}, \citenamefont {Isakov},
  \citenamefont {Smelyanskiy}, \citenamefont {Babbush}, \citenamefont {Ding},
  \citenamefont {Jiang}, \citenamefont {Martinis},\ and\ \citenamefont
  {Neven}}]{Boixo:2016aa}%
  \BibitemOpen
  \bibfield  {author} {\bibinfo {author} {\bibfnamefont {S.}~\bibnamefont
  {Boixo}}, \bibinfo {author} {\bibfnamefont {S.~V.}\ \bibnamefont {Isakov}},
  \bibinfo {author} {\bibfnamefont {V.~N.}\ \bibnamefont {Smelyanskiy}},
  \bibinfo {author} {\bibfnamefont {R.}~\bibnamefont {Babbush}}, \bibinfo
  {author} {\bibfnamefont {N.}~\bibnamefont {Ding}}, \bibinfo {author}
  {\bibfnamefont {Z.}~\bibnamefont {Jiang}}, \bibinfo {author} {\bibfnamefont
  {J.~M.}\ \bibnamefont {Martinis}}, \ and\ \bibinfo {author} {\bibfnamefont
  {H.}~\bibnamefont {Neven}},\ }\href {http://arXiv.org/abs/1608.00263}
  {\bibfield  {journal} {\bibinfo  {journal} {arXiv:1608.00263}\ } (\bibinfo
  {year} {2016}{\natexlab{a}})}\BibitemShut {NoStop}%
\bibitem [{\citenamefont {Fefferman}\ \emph {et~al.}(2017)\citenamefont
  {Fefferman}, \citenamefont {Foss-Feig},\ and\ \citenamefont
  {Gorshkov}}]{Fefferman:2017ab}%
  \BibitemOpen
  \bibfield  {author} {\bibinfo {author} {\bibfnamefont {B.}~\bibnamefont
  {Fefferman}}, \bibinfo {author} {\bibfnamefont {M.}~\bibnamefont
  {Foss-Feig}}, \ and\ \bibinfo {author} {\bibfnamefont {A.~V.}\ \bibnamefont
  {Gorshkov}},\ }\href {http://arXiv.org/abs/1701.03167} {\bibfield  {journal}
  {\bibinfo  {journal} {arXiv:1701.03167}\ } (\bibinfo {year}
  {2017})}\BibitemShut {NoStop}%
\bibitem [{\citenamefont {R{\o}nnow}\ \emph {et~al.}(2014)\citenamefont
  {R{\o}nnow}, \citenamefont {Wang}, \citenamefont {Job}, \citenamefont
  {Boixo}, \citenamefont {Isakov}, \citenamefont {Wecker}, \citenamefont
  {Martinis}, \citenamefont {Lidar},\ and\ \citenamefont {Troyer}}]{speedup}%
  \BibitemOpen
  \bibfield  {author} {\bibinfo {author} {\bibfnamefont {T.~F.}\ \bibnamefont
  {R{\o}nnow}}, \bibinfo {author} {\bibfnamefont {Z.}~\bibnamefont {Wang}},
  \bibinfo {author} {\bibfnamefont {J.}~\bibnamefont {Job}}, \bibinfo {author}
  {\bibfnamefont {S.}~\bibnamefont {Boixo}}, \bibinfo {author} {\bibfnamefont
  {S.~V.}\ \bibnamefont {Isakov}}, \bibinfo {author} {\bibfnamefont
  {D.}~\bibnamefont {Wecker}}, \bibinfo {author} {\bibfnamefont {J.~M.}\
  \bibnamefont {Martinis}}, \bibinfo {author} {\bibfnamefont {D.~A.}\
  \bibnamefont {Lidar}}, \ and\ \bibinfo {author} {\bibfnamefont
  {M.}~\bibnamefont {Troyer}},\ }\href
  {http://science.sciencemag.org/content/345/6195/420} {\bibfield  {journal}
  {\bibinfo  {journal} {Science}\ }\textbf {\bibinfo {volume} {345}},\ \bibinfo
  {pages} {420} (\bibinfo {year} {2014})}\BibitemShut {NoStop}%
\bibitem [{\citenamefont {Shor}(1997)}]{Shor:97}%
  \BibitemOpen
  \bibfield  {author} {\bibinfo {author} {\bibfnamefont {P.~W.}\ \bibnamefont
  {Shor}},\ }\href@noop {} {\bibfield  {journal} {\bibinfo  {journal} {SIAM J.
  Comput.}\ }\textbf {\bibinfo {volume} {26}},\ \bibinfo {pages} {1484}
  (\bibinfo {year} {1997})}\BibitemShut {NoStop}%
\bibitem [{\citenamefont {Farhi}\ \emph {et~al.}(2014)\citenamefont {Farhi},
  \citenamefont {Goldstone},\ and\ \citenamefont {Gutmann}}]{Farhi:2014aa}%
  \BibitemOpen
  \bibfield  {author} {\bibinfo {author} {\bibfnamefont {E.}~\bibnamefont
  {Farhi}}, \bibinfo {author} {\bibfnamefont {J.}~\bibnamefont {Goldstone}}, \
  and\ \bibinfo {author} {\bibfnamefont {S.}~\bibnamefont {Gutmann}},\ }\href
  {http://arXiv.org/abs/1412.6062} {\bibfield  {journal} {\bibinfo  {journal}
  {arXiv:1412.6062}\ } (\bibinfo {year} {2014})}\BibitemShut {NoStop}%
\bibitem [{\citenamefont {Yip}\ \emph {et~al.}(2017)\citenamefont {Yip},
  \citenamefont {Albash},\ and\ \citenamefont {Lidar}}]{Yip:2017}%
  \BibitemOpen
  \bibfield  {author} {\bibinfo {author} {\bibfnamefont {K.-W.}\ \bibnamefont
  {Yip}}, \bibinfo {author} {\bibfnamefont {T.}~\bibnamefont {Albash}}, \ and\
  \bibinfo {author} {\bibfnamefont {D.}~\bibnamefont {Lidar}},\ }\href@noop {}
  {\enquote {\bibinfo {title} {Quantum trajectories for time-dependent
  adiabatic master equations},}\ }\bibinfo {howpublished} {in preparation}
  (\bibinfo {year} {2017})\BibitemShut {NoStop}%
\bibitem [{\citenamefont {Lanting}\ \emph {et~al.}(2014)\citenamefont
  {Lanting}, \citenamefont {Przybysz}, \citenamefont {Smirnov}, \citenamefont
  {Spedalieri}, \citenamefont {Amin}, \citenamefont {Berkley}, \citenamefont
  {Harris}, \citenamefont {Altomare}, \citenamefont {Boixo}, \citenamefont
  {Bunyk}, \citenamefont {Dickson}, \citenamefont {Enderud}, \citenamefont
  {Hilton}, \citenamefont {Hoskinson}, \citenamefont {Johnson}, \citenamefont
  {Ladizinsky}, \citenamefont {Ladizinsky}, \citenamefont {Neufeld},
  \citenamefont {Oh}, \citenamefont {Perminov}, \citenamefont {Rich},
  \citenamefont {Thom}, \citenamefont {Tolkacheva}, \citenamefont {Uchaikin},
  \citenamefont {Wilson},\ and\ \citenamefont {Rose}}]{DWave-entanglement}%
  \BibitemOpen
  \bibfield  {author} {\bibinfo {author} {\bibfnamefont {T.}~\bibnamefont
  {Lanting}}, \bibinfo {author} {\bibfnamefont {A.~J.}\ \bibnamefont
  {Przybysz}}, \bibinfo {author} {\bibfnamefont {A.~Y.}\ \bibnamefont
  {Smirnov}}, \bibinfo {author} {\bibfnamefont {F.~M.}\ \bibnamefont
  {Spedalieri}}, \bibinfo {author} {\bibfnamefont {M.~H.}\ \bibnamefont
  {Amin}}, \bibinfo {author} {\bibfnamefont {A.~J.}\ \bibnamefont {Berkley}},
  \bibinfo {author} {\bibfnamefont {R.}~\bibnamefont {Harris}}, \bibinfo
  {author} {\bibfnamefont {F.}~\bibnamefont {Altomare}}, \bibinfo {author}
  {\bibfnamefont {S.}~\bibnamefont {Boixo}}, \bibinfo {author} {\bibfnamefont
  {P.}~\bibnamefont {Bunyk}}, \bibinfo {author} {\bibfnamefont
  {N.}~\bibnamefont {Dickson}}, \bibinfo {author} {\bibfnamefont
  {C.}~\bibnamefont {Enderud}}, \bibinfo {author} {\bibfnamefont {J.~P.}\
  \bibnamefont {Hilton}}, \bibinfo {author} {\bibfnamefont {E.}~\bibnamefont
  {Hoskinson}}, \bibinfo {author} {\bibfnamefont {M.~W.}\ \bibnamefont
  {Johnson}}, \bibinfo {author} {\bibfnamefont {E.}~\bibnamefont {Ladizinsky}},
  \bibinfo {author} {\bibfnamefont {N.}~\bibnamefont {Ladizinsky}}, \bibinfo
  {author} {\bibfnamefont {R.}~\bibnamefont {Neufeld}}, \bibinfo {author}
  {\bibfnamefont {T.}~\bibnamefont {Oh}}, \bibinfo {author} {\bibfnamefont
  {I.}~\bibnamefont {Perminov}}, \bibinfo {author} {\bibfnamefont
  {C.}~\bibnamefont {Rich}}, \bibinfo {author} {\bibfnamefont {M.~C.}\
  \bibnamefont {Thom}}, \bibinfo {author} {\bibfnamefont {E.}~\bibnamefont
  {Tolkacheva}}, \bibinfo {author} {\bibfnamefont {S.}~\bibnamefont
  {Uchaikin}}, \bibinfo {author} {\bibfnamefont {A.~B.}\ \bibnamefont
  {Wilson}}, \ and\ \bibinfo {author} {\bibfnamefont {G.}~\bibnamefont
  {Rose}},\ }\href {\doibase 10.1103/PhysRevX.4.021041} {\bibfield  {journal}
  {\bibinfo  {journal} {Phys. Rev. X}\ }\textbf {\bibinfo {volume} {4}},\
  \bibinfo {pages} {021041} (\bibinfo {year} {2014})}\BibitemShut {NoStop}%
\bibitem [{\citenamefont {Vidal}\ and\ \citenamefont
  {Werner}(2002)}]{Vidal:02a}%
  \BibitemOpen
  \bibfield  {author} {\bibinfo {author} {\bibfnamefont {G.}~\bibnamefont
  {Vidal}}\ and\ \bibinfo {author} {\bibfnamefont {R.~F.}\ \bibnamefont
  {Werner}},\ }\href {http://link.aps.org/doi/10.1103/PhysRevA.65.032314}
  {\bibfield  {journal} {\bibinfo  {journal} {Phys. Rev. A}\ }\textbf {\bibinfo
  {volume} {65}},\ \bibinfo {pages} {032314} (\bibinfo {year}
  {2002})}\BibitemShut {NoStop}%
\bibitem [{\citenamefont {Spedalieri}(2012)}]{Spedalieri:2012fk}%
  \BibitemOpen
  \bibfield  {author} {\bibinfo {author} {\bibfnamefont {F.~M.}\ \bibnamefont
  {Spedalieri}},\ }\href {http://link.aps.org/doi/10.1103/PhysRevA.86.062311}
  {\bibfield  {journal} {\bibinfo  {journal} {Phys. Rev. A}\ }\textbf {\bibinfo
  {volume} {86}},\ \bibinfo {pages} {062311} (\bibinfo {year}
  {2012})}\BibitemShut {NoStop}%
\bibitem [{\citenamefont {Albash}\ \emph
  {et~al.}(2015{\natexlab{a}})\citenamefont {Albash}, \citenamefont {Hen},
  \citenamefont {Spedalieri},\ and\ \citenamefont {Lidar}}]{Albash:2015pd}%
  \BibitemOpen
  \bibfield  {author} {\bibinfo {author} {\bibfnamefont {T.}~\bibnamefont
  {Albash}}, \bibinfo {author} {\bibfnamefont {I.}~\bibnamefont {Hen}},
  \bibinfo {author} {\bibfnamefont {F.~M.}\ \bibnamefont {Spedalieri}}, \ and\
  \bibinfo {author} {\bibfnamefont {D.~A.}\ \bibnamefont {Lidar}},\ }\href
  {http://link.aps.org/doi/10.1103/PhysRevA.92.062328} {\bibfield  {journal}
  {\bibinfo  {journal} {Physical Review A}\ }\textbf {\bibinfo {volume} {92}},\
  \bibinfo {pages} {062328} (\bibinfo {year} {2015}{\natexlab{a}})}\BibitemShut
  {NoStop}%
\bibitem [{\citenamefont {Albash}\ \emph {et~al.}(2012)\citenamefont {Albash},
  \citenamefont {Boixo}, \citenamefont {Lidar},\ and\ \citenamefont
  {Zanardi}}]{aqcME}%
  \BibitemOpen
  \bibfield  {author} {\bibinfo {author} {\bibfnamefont {T.}~\bibnamefont
  {Albash}}, \bibinfo {author} {\bibfnamefont {S.}~\bibnamefont {Boixo}},
  \bibinfo {author} {\bibfnamefont {D.~A.}\ \bibnamefont {Lidar}}, \ and\
  \bibinfo {author} {\bibfnamefont {P.}~\bibnamefont {Zanardi}},\ }\href
  {http://stacks.iop.org/1367-2630/14/i=12/a=123016} {\bibfield  {journal}
  {\bibinfo  {journal} {New Journal of Physics}\ }\textbf {\bibinfo {volume}
  {14}},\ \bibinfo {pages} {123016} (\bibinfo {year} {2012})}\BibitemShut
  {NoStop}%
\bibitem [{\citenamefont {Boixo}\ \emph {et~al.}(2013)\citenamefont {Boixo},
  \citenamefont {Albash}, \citenamefont {Spedalieri}, \citenamefont
  {Chancellor},\ and\ \citenamefont {Lidar}}]{q-sig}%
  \BibitemOpen
  \bibfield  {author} {\bibinfo {author} {\bibfnamefont {S.}~\bibnamefont
  {Boixo}}, \bibinfo {author} {\bibfnamefont {T.}~\bibnamefont {Albash}},
  \bibinfo {author} {\bibfnamefont {F.~M.}\ \bibnamefont {Spedalieri}},
  \bibinfo {author} {\bibfnamefont {N.}~\bibnamefont {Chancellor}}, \ and\
  \bibinfo {author} {\bibfnamefont {D.~A.}\ \bibnamefont {Lidar}},\ }\href
  {\doibase 10.1038/ncomms3067} {\bibfield  {journal} {\bibinfo  {journal}
  {Nat. Commun.}\ }\textbf {\bibinfo {volume} {4}},\ \bibinfo {pages} {2067}
  (\bibinfo {year} {2013})}\BibitemShut {NoStop}%
\bibitem [{\citenamefont {Smolin}\ and\ \citenamefont {Smith}(2014)}]{Smolin}%
  \BibitemOpen
  \bibfield  {author} {\bibinfo {author} {\bibfnamefont {J.~A.}\ \bibnamefont
  {Smolin}}\ and\ \bibinfo {author} {\bibfnamefont {G.}~\bibnamefont {Smith}},\
  }\href {http://journal.frontiersin.org/article/10.3389/fphy.2014.00052}
  {\bibfield  {journal} {\bibinfo  {journal} {Frontiers in Physics}\ }\textbf
  {\bibinfo {volume} {2}},\ \bibinfo {pages} {52} (\bibinfo {year}
  {2014})}\BibitemShut {NoStop}%
\bibitem [{\citenamefont {Wang}\ \emph {et~al.}(2013)\citenamefont {Wang},
  \citenamefont {R{\o}nnow}, \citenamefont {Boixo}, \citenamefont {Isakov},
  \citenamefont {Wang}, \citenamefont {Wecker}, \citenamefont {Lidar},
  \citenamefont {Martinis},\ and\ \citenamefont {Troyer}}]{comment-SS}%
  \BibitemOpen
  \bibfield  {author} {\bibinfo {author} {\bibfnamefont {L.}~\bibnamefont
  {Wang}}, \bibinfo {author} {\bibfnamefont {T.~F.}\ \bibnamefont {R{\o}nnow}},
  \bibinfo {author} {\bibfnamefont {S.}~\bibnamefont {Boixo}}, \bibinfo
  {author} {\bibfnamefont {S.~V.}\ \bibnamefont {Isakov}}, \bibinfo {author}
  {\bibfnamefont {Z.}~\bibnamefont {Wang}}, \bibinfo {author} {\bibfnamefont
  {D.}~\bibnamefont {Wecker}}, \bibinfo {author} {\bibfnamefont {D.~A.}\
  \bibnamefont {Lidar}}, \bibinfo {author} {\bibfnamefont {J.~M.}\ \bibnamefont
  {Martinis}}, \ and\ \bibinfo {author} {\bibfnamefont {M.}~\bibnamefont
  {Troyer}},\ }\href {http://arxiv.org/abs/1305.5837} {\bibfield  {journal}
  {\bibinfo  {journal} {arXiv:1305.5837}\ } (\bibinfo {year}
  {2013})}\BibitemShut {NoStop}%
\bibitem [{\citenamefont {Albash}\ \emph
  {et~al.}(2015{\natexlab{b}})\citenamefont {Albash}, \citenamefont {Vinci},
  \citenamefont {Mishra}, \citenamefont {Warburton},\ and\ \citenamefont
  {Lidar}}]{q-sig2}%
  \BibitemOpen
  \bibfield  {author} {\bibinfo {author} {\bibfnamefont {T.}~\bibnamefont
  {Albash}}, \bibinfo {author} {\bibfnamefont {W.}~\bibnamefont {Vinci}},
  \bibinfo {author} {\bibfnamefont {A.}~\bibnamefont {Mishra}}, \bibinfo
  {author} {\bibfnamefont {P.~A.}\ \bibnamefont {Warburton}}, \ and\ \bibinfo
  {author} {\bibfnamefont {D.~A.}\ \bibnamefont {Lidar}},\ }\href
  {http://link.aps.org/doi/10.1103/PhysRevA.91.042314} {\bibfield  {journal}
  {\bibinfo  {journal} {Phys. Rev. A}\ }\textbf {\bibinfo {volume} {91}},\
  \bibinfo {pages} {042314} (\bibinfo {year} {2015}{\natexlab{b}})}\BibitemShut
  {NoStop}%
\bibitem [{\citenamefont {Shin}\ \emph
  {et~al.}(2014{\natexlab{a}})\citenamefont {Shin}, \citenamefont {Smith},
  \citenamefont {Smolin},\ and\ \citenamefont {Vazirani}}]{SSSV-comment}%
  \BibitemOpen
  \bibfield  {author} {\bibinfo {author} {\bibfnamefont {S.~W.}\ \bibnamefont
  {Shin}}, \bibinfo {author} {\bibfnamefont {G.}~\bibnamefont {Smith}},
  \bibinfo {author} {\bibfnamefont {J.~A.}\ \bibnamefont {Smolin}}, \ and\
  \bibinfo {author} {\bibfnamefont {U.}~\bibnamefont {Vazirani}},\ }\href
  {http://arXiv.org/abs/1404.6499} {\bibfield  {journal} {\bibinfo  {journal}
  {arXiv:1404.6499}\ } (\bibinfo {year} {2014}{\natexlab{a}})}\BibitemShut
  {NoStop}%
\bibitem [{\citenamefont {Kirkpatrick}\ \emph {et~al.}(1983)\citenamefont
  {Kirkpatrick}, \citenamefont {Gelatt},\ and\ \citenamefont
  {Vecchi}}]{kirkpatrick_optimization_1983}%
  \BibitemOpen
  \bibfield  {author} {\bibinfo {author} {\bibfnamefont {S.}~\bibnamefont
  {Kirkpatrick}}, \bibinfo {author} {\bibfnamefont {C.~D.}\ \bibnamefont
  {Gelatt}}, \ and\ \bibinfo {author} {\bibfnamefont {M.~P.}\ \bibnamefont
  {Vecchi}},\ }\href {http://science.sciencemag.org/content/220/4598/671}
  {\bibfield  {journal} {\bibinfo  {journal} {Science}\ }\textbf {\bibinfo
  {volume} {220}},\ \bibinfo {pages} {671} (\bibinfo {year}
  {1983})}\BibitemShut {NoStop}%
\bibitem [{\citenamefont {Boixo}\ \emph {et~al.}(2014)\citenamefont {Boixo},
  \citenamefont {Ronnow}, \citenamefont {Isakov}, \citenamefont {Wang},
  \citenamefont {Wecker}, \citenamefont {Lidar}, \citenamefont {Martinis},\
  and\ \citenamefont {Troyer}}]{q108}%
  \BibitemOpen
  \bibfield  {author} {\bibinfo {author} {\bibfnamefont {S.}~\bibnamefont
  {Boixo}}, \bibinfo {author} {\bibfnamefont {T.~F.}\ \bibnamefont {Ronnow}},
  \bibinfo {author} {\bibfnamefont {S.~V.}\ \bibnamefont {Isakov}}, \bibinfo
  {author} {\bibfnamefont {Z.}~\bibnamefont {Wang}}, \bibinfo {author}
  {\bibfnamefont {D.}~\bibnamefont {Wecker}}, \bibinfo {author} {\bibfnamefont
  {D.~A.}\ \bibnamefont {Lidar}}, \bibinfo {author} {\bibfnamefont {J.~M.}\
  \bibnamefont {Martinis}}, \ and\ \bibinfo {author} {\bibfnamefont
  {M.}~\bibnamefont {Troyer}},\ }\href {\doibase 10.1038/nphys2900} {\bibfield
  {journal} {\bibinfo  {journal} {Nat. Phys.}\ }\textbf {\bibinfo {volume}
  {10}},\ \bibinfo {pages} {218} (\bibinfo {year} {2014})}\BibitemShut
  {NoStop}%
\bibitem [{\citenamefont {Marto\ifmmode~\check{n}\else \v{n}\fi{}\'ak}\ \emph
  {et~al.}(2002)\citenamefont {Marto\ifmmode~\check{n}\else \v{n}\fi{}\'ak},
  \citenamefont {Santoro},\ and\ \citenamefont {Tosatti}}]{sqa1}%
  \BibitemOpen
  \bibfield  {author} {\bibinfo {author} {\bibfnamefont {R.}~\bibnamefont
  {Marto\ifmmode~\check{n}\else \v{n}\fi{}\'ak}}, \bibinfo {author}
  {\bibfnamefont {G.~E.}\ \bibnamefont {Santoro}}, \ and\ \bibinfo {author}
  {\bibfnamefont {E.}~\bibnamefont {Tosatti}},\ }\href {\doibase
  10.1103/PhysRevB.66.094203} {\bibfield  {journal} {\bibinfo  {journal} {Phys.
  Rev. B}\ }\textbf {\bibinfo {volume} {66}},\ \bibinfo {pages} {094203}
  (\bibinfo {year} {2002})}\BibitemShut {NoStop}%
\bibitem [{\citenamefont {Shin}\ \emph
  {et~al.}(2014{\natexlab{b}})\citenamefont {Shin}, \citenamefont {Smith},
  \citenamefont {Smolin},\ and\ \citenamefont {Vazirani}}]{SSSV}%
  \BibitemOpen
  \bibfield  {author} {\bibinfo {author} {\bibfnamefont {S.~W.}\ \bibnamefont
  {Shin}}, \bibinfo {author} {\bibfnamefont {G.}~\bibnamefont {Smith}},
  \bibinfo {author} {\bibfnamefont {J.~A.}\ \bibnamefont {Smolin}}, \ and\
  \bibinfo {author} {\bibfnamefont {U.}~\bibnamefont {Vazirani}},\ }\href
  {http://arXiv.org/abs/1401.7087} {\bibfield  {journal} {\bibinfo  {journal}
  {arXiv:1401.7087}\ } (\bibinfo {year} {2014}{\natexlab{b}})}\BibitemShut
  {NoStop}%
\bibitem [{\citenamefont {Crowley}\ and\ \citenamefont
  {Green}(2016)}]{Crowley:2016aa}%
  \BibitemOpen
  \bibfield  {author} {\bibinfo {author} {\bibfnamefont {P.~J.~D.}\
  \bibnamefont {Crowley}}\ and\ \bibinfo {author} {\bibfnamefont {A.~G.}\
  \bibnamefont {Green}},\ }\href
  {https://link.aps.org/doi/10.1103/PhysRevA.94.062106} {\bibfield  {journal}
  {\bibinfo  {journal} {Physical Review A}\ }\textbf {\bibinfo {volume} {94}},\
  \bibinfo {pages} {062106} (\bibinfo {year} {2016})}\BibitemShut {NoStop}%
\bibitem [{\citenamefont {Albash}\ \emph
  {et~al.}(2015{\natexlab{c}})\citenamefont {Albash}, \citenamefont
  {R{\o}nnow}, \citenamefont {Troyer},\ and\ \citenamefont
  {Lidar}}]{Albash:2014if}%
  \BibitemOpen
  \bibfield  {author} {\bibinfo {author} {\bibfnamefont {T.}~\bibnamefont
  {Albash}}, \bibinfo {author} {\bibfnamefont {T.~F.}\ \bibnamefont
  {R{\o}nnow}}, \bibinfo {author} {\bibfnamefont {M.}~\bibnamefont {Troyer}}, \
  and\ \bibinfo {author} {\bibfnamefont {D.~A.}\ \bibnamefont {Lidar}},\ }\href
  {https://link.springer.com/article/10.1140%2Fepjst%2Fe2015-02346-0}
  {\bibfield  {journal} {\bibinfo  {journal} {Eur. Phys. J. Spec. Top.}\
  }\textbf {\bibinfo {volume} {224}},\ \bibinfo {pages} {111} (\bibinfo {year}
  {2015}{\natexlab{c}})}\BibitemShut {NoStop}%
\bibitem [{\citenamefont {Brooke}\ \emph {et~al.}(1999)\citenamefont {Brooke},
  \citenamefont {Bitko}, \citenamefont {F.}, \citenamefont {Rosenbaum},\ and\
  \citenamefont {Aeppli}}]{Brooke1999}%
  \BibitemOpen
  \bibfield  {author} {\bibinfo {author} {\bibfnamefont {J.}~\bibnamefont
  {Brooke}}, \bibinfo {author} {\bibfnamefont {D.}~\bibnamefont {Bitko}},
  \bibinfo {author} {\bibfnamefont {T.}~\bibnamefont {F.}}, \bibinfo {author}
  {\bibnamefont {Rosenbaum}}, \ and\ \bibinfo {author} {\bibfnamefont
  {G.}~\bibnamefont {Aeppli}},\ }\href {\doibase 10.1126/science.284.5415.779}
  {\bibfield  {journal} {\bibinfo  {journal} {Science}\ }\textbf {\bibinfo
  {volume} {284}},\ \bibinfo {pages} {779} (\bibinfo {year}
  {1999})}\BibitemShut {NoStop}%
\bibitem [{\citenamefont {Brooke}\ \emph {et~al.}(2001)\citenamefont {Brooke},
  \citenamefont {Rosenbaum},\ and\ \citenamefont
  {Aeppli}}]{brooke_tunable_2001}%
  \BibitemOpen
  \bibfield  {author} {\bibinfo {author} {\bibfnamefont {J.}~\bibnamefont
  {Brooke}}, \bibinfo {author} {\bibfnamefont {T.~F.}\ \bibnamefont
  {Rosenbaum}}, \ and\ \bibinfo {author} {\bibfnamefont {G.}~\bibnamefont
  {Aeppli}},\ }\href
  {http://www.nature.com/nature/journal/v413/n6856/full/413610a0.html}
  {\bibfield  {journal} {\bibinfo  {journal} {Nature}\ }\textbf {\bibinfo
  {volume} {413}},\ \bibinfo {pages} {610} (\bibinfo {year}
  {2001})}\BibitemShut {NoStop}%
\bibitem [{\citenamefont {Johnson}\ \emph {et~al.}(2011)\citenamefont
  {Johnson}, \citenamefont {Amin}, \citenamefont {Gildert}, \citenamefont
  {Lanting}, \citenamefont {Hamze}, \citenamefont {Dickson}, \citenamefont
  {Harris}, \citenamefont {Berkley}, \citenamefont {Johansson}, \citenamefont
  {Bunyk}, \citenamefont {Chapple}, \citenamefont {Enderud}, \citenamefont
  {Hilton}, \citenamefont {Karimi}, \citenamefont {Ladizinsky}, \citenamefont
  {Ladizinsky}, \citenamefont {Oh}, \citenamefont {Perminov}, \citenamefont
  {Rich}, \citenamefont {Thom}, \citenamefont {Tolkacheva}, \citenamefont
  {Truncik}, \citenamefont {Uchaikin}, \citenamefont {Wang}, \citenamefont
  {Wilson},\ and\ \citenamefont {Rose}}]{DWave}%
  \BibitemOpen
  \bibfield  {author} {\bibinfo {author} {\bibfnamefont {M.~W.}\ \bibnamefont
  {Johnson}}, \bibinfo {author} {\bibfnamefont {M.~H.~S.}\ \bibnamefont
  {Amin}}, \bibinfo {author} {\bibfnamefont {S.}~\bibnamefont {Gildert}},
  \bibinfo {author} {\bibfnamefont {T.}~\bibnamefont {Lanting}}, \bibinfo
  {author} {\bibfnamefont {F.}~\bibnamefont {Hamze}}, \bibinfo {author}
  {\bibfnamefont {N.}~\bibnamefont {Dickson}}, \bibinfo {author} {\bibfnamefont
  {R.}~\bibnamefont {Harris}}, \bibinfo {author} {\bibfnamefont {A.~J.}\
  \bibnamefont {Berkley}}, \bibinfo {author} {\bibfnamefont {J.}~\bibnamefont
  {Johansson}}, \bibinfo {author} {\bibfnamefont {P.}~\bibnamefont {Bunyk}},
  \bibinfo {author} {\bibfnamefont {E.~M.}\ \bibnamefont {Chapple}}, \bibinfo
  {author} {\bibfnamefont {C.}~\bibnamefont {Enderud}}, \bibinfo {author}
  {\bibfnamefont {J.~P.}\ \bibnamefont {Hilton}}, \bibinfo {author}
  {\bibfnamefont {K.}~\bibnamefont {Karimi}}, \bibinfo {author} {\bibfnamefont
  {E.}~\bibnamefont {Ladizinsky}}, \bibinfo {author} {\bibfnamefont
  {N.}~\bibnamefont {Ladizinsky}}, \bibinfo {author} {\bibfnamefont
  {T.}~\bibnamefont {Oh}}, \bibinfo {author} {\bibfnamefont {I.}~\bibnamefont
  {Perminov}}, \bibinfo {author} {\bibfnamefont {C.}~\bibnamefont {Rich}},
  \bibinfo {author} {\bibfnamefont {M.~C.}\ \bibnamefont {Thom}}, \bibinfo
  {author} {\bibfnamefont {E.}~\bibnamefont {Tolkacheva}}, \bibinfo {author}
  {\bibfnamefont {C.~J.~S.}\ \bibnamefont {Truncik}}, \bibinfo {author}
  {\bibfnamefont {S.}~\bibnamefont {Uchaikin}}, \bibinfo {author}
  {\bibfnamefont {J.}~\bibnamefont {Wang}}, \bibinfo {author} {\bibfnamefont
  {B.}~\bibnamefont {Wilson}}, \ and\ \bibinfo {author} {\bibfnamefont
  {G.}~\bibnamefont {Rose}},\ }\href
  {https://www.nature.com/nature/journal/v473/n7346/full/nature10012.html}
  {\bibfield  {journal} {\bibinfo  {journal} {Nature}\ }\textbf {\bibinfo
  {volume} {473}},\ \bibinfo {pages} {194} (\bibinfo {year}
  {2011})}\BibitemShut {NoStop}%
\bibitem [{\citenamefont {Boixo}\ \emph
  {et~al.}(2016{\natexlab{b}})\citenamefont {Boixo}, \citenamefont
  {Smelyanskiy}, \citenamefont {Shabani}, \citenamefont {Isakov}, \citenamefont
  {Dykman}, \citenamefont {Denchev}, \citenamefont {Amin}, \citenamefont
  {Smirnov}, \citenamefont {Mohseni},\ and\ \citenamefont
  {Neven}}]{Boixo:2014yu}%
  \BibitemOpen
  \bibfield  {author} {\bibinfo {author} {\bibfnamefont {S.}~\bibnamefont
  {Boixo}}, \bibinfo {author} {\bibfnamefont {V.~N.}\ \bibnamefont
  {Smelyanskiy}}, \bibinfo {author} {\bibfnamefont {A.}~\bibnamefont
  {Shabani}}, \bibinfo {author} {\bibfnamefont {S.~V.}\ \bibnamefont {Isakov}},
  \bibinfo {author} {\bibfnamefont {M.}~\bibnamefont {Dykman}}, \bibinfo
  {author} {\bibfnamefont {V.~S.}\ \bibnamefont {Denchev}}, \bibinfo {author}
  {\bibfnamefont {M.~H.}\ \bibnamefont {Amin}}, \bibinfo {author}
  {\bibfnamefont {A.~Y.}\ \bibnamefont {Smirnov}}, \bibinfo {author}
  {\bibfnamefont {M.}~\bibnamefont {Mohseni}}, \ and\ \bibinfo {author}
  {\bibfnamefont {H.}~\bibnamefont {Neven}},\ }\href
  {http://dx.doi.org/10.1038/ncomms10327} {\bibfield  {journal} {\bibinfo
  {journal} {Nat Commun}\ }\textbf {\bibinfo {volume} {7}} (\bibinfo {year}
  {2016}{\natexlab{b}})}\BibitemShut {NoStop}%
\bibitem [{\citenamefont {Denchev}\ \emph {et~al.}(2016)\citenamefont
  {Denchev}, \citenamefont {Boixo}, \citenamefont {Isakov}, \citenamefont
  {Ding}, \citenamefont {Babbush}, \citenamefont {Smelyanskiy}, \citenamefont
  {Martinis},\ and\ \citenamefont {Neven}}]{PhysRevX.6.031015}%
  \BibitemOpen
  \bibfield  {author} {\bibinfo {author} {\bibfnamefont {V.~S.}\ \bibnamefont
  {Denchev}}, \bibinfo {author} {\bibfnamefont {S.}~\bibnamefont {Boixo}},
  \bibinfo {author} {\bibfnamefont {S.~V.}\ \bibnamefont {Isakov}}, \bibinfo
  {author} {\bibfnamefont {N.}~\bibnamefont {Ding}}, \bibinfo {author}
  {\bibfnamefont {R.}~\bibnamefont {Babbush}}, \bibinfo {author} {\bibfnamefont
  {V.}~\bibnamefont {Smelyanskiy}}, \bibinfo {author} {\bibfnamefont
  {J.}~\bibnamefont {Martinis}}, \ and\ \bibinfo {author} {\bibfnamefont
  {H.}~\bibnamefont {Neven}},\ }\href
  {http://link.aps.org/doi/10.1103/PhysRevX.6.031015} {\bibfield  {journal}
  {\bibinfo  {journal} {Phys. Rev. X}\ }\textbf {\bibinfo {volume} {6}},\
  \bibinfo {pages} {031015} (\bibinfo {year} {2016})}\BibitemShut {NoStop}%
\bibitem [{\citenamefont {Mandr{\`a}}\ \emph {et~al.}(2016)\citenamefont
  {Mandr{\`a}}, \citenamefont {Zhu}, \citenamefont {Wang}, \citenamefont
  {Perdomo-Ortiz},\ and\ \citenamefont {Katzgraber}}]{2016arXiv160401746M}%
  \BibitemOpen
  \bibfield  {author} {\bibinfo {author} {\bibfnamefont {S.}~\bibnamefont
  {Mandr{\`a}}}, \bibinfo {author} {\bibfnamefont {Z.}~\bibnamefont {Zhu}},
  \bibinfo {author} {\bibfnamefont {W.}~\bibnamefont {Wang}}, \bibinfo {author}
  {\bibfnamefont {A.}~\bibnamefont {Perdomo-Ortiz}}, \ and\ \bibinfo {author}
  {\bibfnamefont {H.~G.}\ \bibnamefont {Katzgraber}},\ }\href
  {http://link.aps.org/doi/10.1103/PhysRevA.94.022337} {\bibfield  {journal}
  {\bibinfo  {journal} {Physical Review A}\ }\textbf {\bibinfo {volume} {94}},\
  \bibinfo {pages} {022337} (\bibinfo {year} {2016})}\BibitemShut {NoStop}%
\bibitem [{\citenamefont {Monz}\ \emph {et~al.}(2011)\citenamefont {Monz},
  \citenamefont {Schindler}, \citenamefont {Barreiro}, \citenamefont {Chwalla},
  \citenamefont {Nigg}, \citenamefont {Coish}, \citenamefont {Harlander},
  \citenamefont {H\"ansel}, \citenamefont {Hennrich},\ and\ \citenamefont
  {Blatt}}]{PhysRevLett.106.130506}%
  \BibitemOpen
  \bibfield  {author} {\bibinfo {author} {\bibfnamefont {T.}~\bibnamefont
  {Monz}}, \bibinfo {author} {\bibfnamefont {P.}~\bibnamefont {Schindler}},
  \bibinfo {author} {\bibfnamefont {J.~T.}\ \bibnamefont {Barreiro}}, \bibinfo
  {author} {\bibfnamefont {M.}~\bibnamefont {Chwalla}}, \bibinfo {author}
  {\bibfnamefont {D.}~\bibnamefont {Nigg}}, \bibinfo {author} {\bibfnamefont
  {W.~A.}\ \bibnamefont {Coish}}, \bibinfo {author} {\bibfnamefont
  {M.}~\bibnamefont {Harlander}}, \bibinfo {author} {\bibfnamefont
  {W.}~\bibnamefont {H\"ansel}}, \bibinfo {author} {\bibfnamefont
  {M.}~\bibnamefont {Hennrich}}, \ and\ \bibinfo {author} {\bibfnamefont
  {R.}~\bibnamefont {Blatt}},\ }\href {\doibase 10.1103/PhysRevLett.106.130506}
  {\bibfield  {journal} {\bibinfo  {journal} {Phys. Rev. Lett.}\ }\textbf
  {\bibinfo {volume} {106}},\ \bibinfo {pages} {130506} (\bibinfo {year}
  {2011})}\BibitemShut {NoStop}%
\bibitem [{\citenamefont {Bohnet}\ \emph {et~al.}(2016)\citenamefont {Bohnet},
  \citenamefont {Sawyer}, \citenamefont {Britton}, \citenamefont {Wall},
  \citenamefont {Rey}, \citenamefont {Foss-Feig},\ and\ \citenamefont
  {Bollinger}}]{Bohnet:2016aa}%
  \BibitemOpen
  \bibfield  {author} {\bibinfo {author} {\bibfnamefont {J.~G.}\ \bibnamefont
  {Bohnet}}, \bibinfo {author} {\bibfnamefont {B.~C.}\ \bibnamefont {Sawyer}},
  \bibinfo {author} {\bibfnamefont {J.~W.}\ \bibnamefont {Britton}}, \bibinfo
  {author} {\bibfnamefont {M.~L.}\ \bibnamefont {Wall}}, \bibinfo {author}
  {\bibfnamefont {A.~M.}\ \bibnamefont {Rey}}, \bibinfo {author} {\bibfnamefont
  {M.}~\bibnamefont {Foss-Feig}}, \ and\ \bibinfo {author} {\bibfnamefont
  {J.~J.}\ \bibnamefont {Bollinger}},\ }\href
  {http://science.sciencemag.org/content/352/6291/1297.abstract} {\bibfield
  {journal} {\bibinfo  {journal} {Science}\ }\textbf {\bibinfo {volume}
  {352}},\ \bibinfo {pages} {1297} (\bibinfo {year} {2016})}\BibitemShut
  {NoStop}%
\bibitem [{\citenamefont {{Catherine C. McGeoch}}(2012)}]{McGeoch:book}%
  \BibitemOpen
  \bibfield  {author} {\bibinfo {author} {\bibnamefont {{Catherine C.
  McGeoch}}},\ }\href
  {http://www.cambridge.org/us/academic/subjects/computer-science/algorithmics-complexity-computer-algebra-and-computational-g/guide-experimental-algorithmics?format=PB&isbn=9780521173018#contentsTabAnchor}
  {\emph {\bibinfo {title} {{A Guide to Experimental Algorithmics}}}}\
  (\bibinfo  {publisher} {{Cambridge University Press}},\ \bibinfo {address}
  {{Cambride, UK}},\ \bibinfo {year} {2012})\BibitemShut {NoStop}%
\bibitem [{\citenamefont {King}\ \emph
  {et~al.}(2015{\natexlab{a}})\citenamefont {King}, \citenamefont {Yarkoni},
  \citenamefont {Nevisi}, \citenamefont {Hilton},\ and\ \citenamefont
  {McGeoch}}]{King:2015cs}%
  \BibitemOpen
  \bibfield  {author} {\bibinfo {author} {\bibfnamefont {J.}~\bibnamefont
  {King}}, \bibinfo {author} {\bibfnamefont {S.}~\bibnamefont {Yarkoni}},
  \bibinfo {author} {\bibfnamefont {M.~M.}\ \bibnamefont {Nevisi}}, \bibinfo
  {author} {\bibfnamefont {J.~P.}\ \bibnamefont {Hilton}}, \ and\ \bibinfo
  {author} {\bibfnamefont {C.~C.}\ \bibnamefont {McGeoch}},\ }\href
  {http://arXiv.org/abs/1508.05087} {\bibfield  {journal} {\bibinfo  {journal}
  {arXiv:1508.05087}\ } (\bibinfo {year} {2015}{\natexlab{a}})}\BibitemShut
  {NoStop}%
\bibitem [{\citenamefont {Vinci}\ and\ \citenamefont
  {Lidar}(2016)}]{Vinci:2016tg}%
  \BibitemOpen
  \bibfield  {author} {\bibinfo {author} {\bibfnamefont {W.}~\bibnamefont
  {Vinci}}\ and\ \bibinfo {author} {\bibfnamefont {D.~A.}\ \bibnamefont
  {Lidar}},\ }\href {http://link.aps.org/doi/10.1103/PhysRevApplied.6.054016}
  {\bibfield  {journal} {\bibinfo  {journal} {Physical Review Applied}\
  }\textbf {\bibinfo {volume} {6}},\ \bibinfo {pages} {054016} (\bibinfo {year}
  {2016})}\BibitemShut {NoStop}%
\bibitem [{\citenamefont {McGeoch}\ and\ \citenamefont {Wang}(2013)}]{McGeoch}%
  \BibitemOpen
  \bibfield  {author} {\bibinfo {author} {\bibfnamefont {C.~C.}\ \bibnamefont
  {McGeoch}}\ and\ \bibinfo {author} {\bibfnamefont {C.}~\bibnamefont {Wang}},\
  }in\ \href@noop {} {\emph {\bibinfo {booktitle} {Proceedings of the 2013 ACM
  Conference on Computing Frontiers}}}\ (\bibinfo {year} {2013})\BibitemShut
  {NoStop}%
\bibitem [{\citenamefont {Santra}\ \emph {et~al.}(2014)\citenamefont {Santra},
  \citenamefont {Quiroz}, \citenamefont {Steeg},\ and\ \citenamefont
  {Lidar}}]{MAX2SAT}%
  \BibitemOpen
  \bibfield  {author} {\bibinfo {author} {\bibfnamefont {S.}~\bibnamefont
  {Santra}}, \bibinfo {author} {\bibfnamefont {G.}~\bibnamefont {Quiroz}},
  \bibinfo {author} {\bibfnamefont {G.~V.}\ \bibnamefont {Steeg}}, \ and\
  \bibinfo {author} {\bibfnamefont {D.~A.}\ \bibnamefont {Lidar}},\ }\href
  {http://stacks.iop.org/1367-2630/16/i=4/a=045006} {\bibfield  {journal}
  {\bibinfo  {journal} {New J. of Phys.}\ }\textbf {\bibinfo {volume} {16}},\
  \bibinfo {pages} {045006} (\bibinfo {year} {2014})}\BibitemShut {NoStop}%
\bibitem [{\citenamefont {Isakov}\ \emph {et~al.}(2015)\citenamefont {Isakov},
  \citenamefont {Zintchenko}, \citenamefont {R{\o}nnow},\ and\ \citenamefont
  {Troyer}}]{Isakov:2015ao}%
  \BibitemOpen
  \bibfield  {author} {\bibinfo {author} {\bibfnamefont {S.~V.}\ \bibnamefont
  {Isakov}}, \bibinfo {author} {\bibfnamefont {I.~N.}\ \bibnamefont
  {Zintchenko}}, \bibinfo {author} {\bibfnamefont {T.~F.}\ \bibnamefont
  {R{\o}nnow}}, \ and\ \bibinfo {author} {\bibfnamefont {M.}~\bibnamefont
  {Troyer}},\ }\href {\doibase http://dx.doi.org/10.1016/j.cpc.2015.02.015}
  {\bibfield  {journal} {\bibinfo  {journal} {Computer Physics Communications}\
  }\textbf {\bibinfo {volume} {192}},\ \bibinfo {pages} {265} (\bibinfo {year}
  {2015})}\BibitemShut {NoStop}%
\bibitem [{\citenamefont {Geyer}(1991)}]{Geyer:91}%
  \BibitemOpen
  \bibfield  {author} {\bibinfo {author} {\bibfnamefont {C.~J.}\ \bibnamefont
  {Geyer}},\ }in\ \href@noop {} {\emph {\bibinfo {booktitle} {Computing Science
  and Statistics Proceedings of the 23rd Symposium on the Interface}}},\
  \bibinfo {editor} {edited by\ \bibinfo {editor} {\bibfnamefont {E.~M.}\
  \bibnamefont {Keramidas}}}\ (\bibinfo  {publisher} {American Statistical
  Association},\ \bibinfo {address} {New York},\ \bibinfo {year} {1991})\ p.\
  \bibinfo {pages} {156}\BibitemShut {NoStop}%
\bibitem [{\citenamefont {Earl}\ and\ \citenamefont
  {Deem}(2005)}]{Earl:2005pd}%
  \BibitemOpen
  \bibfield  {author} {\bibinfo {author} {\bibfnamefont {D.~J.}\ \bibnamefont
  {Earl}}\ and\ \bibinfo {author} {\bibfnamefont {M.~W.}\ \bibnamefont
  {Deem}},\ }\href {\doibase 10.1039/B509983H} {\bibfield  {journal} {\bibinfo
  {journal} {Physical Chemistry Chemical Physics}\ }\textbf {\bibinfo {volume}
  {7}},\ \bibinfo {pages} {3910} (\bibinfo {year} {2005})}\BibitemShut
  {NoStop}%
\bibitem [{\citenamefont {Katzgraber}\ \emph {et~al.}(2006)\citenamefont
  {Katzgraber}, \citenamefont {Trebst}, \citenamefont {Huse},\ and\
  \citenamefont {Troyer}}]{katzgraber:06a}%
  \BibitemOpen
  \bibfield  {author} {\bibinfo {author} {\bibfnamefont {H.~G.}\ \bibnamefont
  {Katzgraber}}, \bibinfo {author} {\bibfnamefont {S.}~\bibnamefont {Trebst}},
  \bibinfo {author} {\bibfnamefont {D.~A.}\ \bibnamefont {Huse}}, \ and\
  \bibinfo {author} {\bibfnamefont {M.}~\bibnamefont {Troyer}},\ }\href
  {\doibase 10.1088/1742-5468/2006/03/P03018} {\bibfield  {journal} {\bibinfo
  {journal} {J. Stat. Mech.}\ }\textbf {\bibinfo {volume} {2006}},\ \bibinfo
  {pages} {P03018} (\bibinfo {year} {2006})}\BibitemShut {NoStop}%
\bibitem [{\citenamefont {Hamze}\ and\ \citenamefont
  {de~Freitas}(2004)}]{hamze:04}%
  \BibitemOpen
  \bibfield  {author} {\bibinfo {author} {\bibfnamefont {F.}~\bibnamefont
  {Hamze}}\ and\ \bibinfo {author} {\bibfnamefont {N.}~\bibnamefont
  {de~Freitas}},\ }in\ \href {http://dl.acm.org/citation.cfm?id=1036873} {\emph
  {\bibinfo {booktitle} {UAI}}},\ \bibinfo {editor} {edited by\ \bibinfo
  {editor} {\bibfnamefont {D.~M.}\ \bibnamefont {Chickering}}\ and\ \bibinfo
  {editor} {\bibfnamefont {J.~Y.}\ \bibnamefont {Halpern}}}\ (\bibinfo
  {publisher} {AUAI Press},\ \bibinfo {address} {Arlington, Virginia},\
  \bibinfo {year} {2004})\ pp.\ \bibinfo {pages} {243--250}\BibitemShut
  {NoStop}%
\bibitem [{\citenamefont {Selby}(2014)}]{Selby:2014tx}%
  \BibitemOpen
  \bibfield  {author} {\bibinfo {author} {\bibfnamefont {A.}~\bibnamefont
  {Selby}},\ }\href {http://arXiv.org/abs/1409.3934} {\bibfield  {journal}
  {\bibinfo  {journal} {arXiv:1409.3934}\ } (\bibinfo {year}
  {2014})}\BibitemShut {NoStop}%
\bibitem [{\citenamefont {Heim}\ \emph {et~al.}(2015)\citenamefont {Heim},
  \citenamefont {R{\o}nnow}, \citenamefont {Isakov},\ and\ \citenamefont
  {Troyer}}]{Heim:2014jf}%
  \BibitemOpen
  \bibfield  {author} {\bibinfo {author} {\bibfnamefont {B.}~\bibnamefont
  {Heim}}, \bibinfo {author} {\bibfnamefont {T.~F.}\ \bibnamefont {R{\o}nnow}},
  \bibinfo {author} {\bibfnamefont {S.~V.}\ \bibnamefont {Isakov}}, \ and\
  \bibinfo {author} {\bibfnamefont {M.}~\bibnamefont {Troyer}},\ }\href
  {http://www.sciencemag.org/content/348/6231/215.abstract} {\bibfield
  {journal} {\bibinfo  {journal} {Science}\ }\textbf {\bibinfo {volume}
  {348}},\ \bibinfo {pages} {215} (\bibinfo {year} {2015})}\BibitemShut
  {NoStop}%
\bibitem [{\citenamefont {Crosson}\ and\ \citenamefont
  {Harrow}(2016)}]{Crosson:2016fk}%
  \BibitemOpen
  \bibfield  {author} {\bibinfo {author} {\bibfnamefont {E.}~\bibnamefont
  {Crosson}}\ and\ \bibinfo {author} {\bibfnamefont {A.~W.}\ \bibnamefont
  {Harrow}},\ }\href {http://arXiv.org/abs/1601.03030} {\bibfield  {journal}
  {\bibinfo  {journal} {arXiv:1601.03030}\ } (\bibinfo {year}
  {2016})}\BibitemShut {NoStop}%
\bibitem [{\citenamefont {Zick}\ \emph {et~al.}(2015)\citenamefont {Zick},
  \citenamefont {Shehab},\ and\ \citenamefont {French}}]{Zick:2015aa}%
  \BibitemOpen
  \bibfield  {author} {\bibinfo {author} {\bibfnamefont {K.~M.}\ \bibnamefont
  {Zick}}, \bibinfo {author} {\bibfnamefont {O.}~\bibnamefont {Shehab}}, \ and\
  \bibinfo {author} {\bibfnamefont {M.}~\bibnamefont {French}},\ }\href
  {https://arxiv.org/ct?url=http%3A%2F%2Fdx.doi.org%2F10%252E1038%2Fsrep11168&v=b092df32}
  {\bibfield  {journal} {\bibinfo  {journal} {arXiv:1503.06453}\ } (\bibinfo
  {year} {2015})}\BibitemShut {NoStop}%
\bibitem [{\citenamefont {Venturelli}\ \emph
  {et~al.}(2015{\natexlab{a}})\citenamefont {Venturelli}, \citenamefont
  {Marchand},\ and\ \citenamefont {Rojo}}]{Venturelli:2015pi}%
  \BibitemOpen
  \bibfield  {author} {\bibinfo {author} {\bibfnamefont {D.}~\bibnamefont
  {Venturelli}}, \bibinfo {author} {\bibfnamefont {D.~J.~J.}\ \bibnamefont
  {Marchand}}, \ and\ \bibinfo {author} {\bibfnamefont {G.}~\bibnamefont
  {Rojo}},\ }\href {http://arXiv.org/abs/1506.08479} {\bibfield  {journal}
  {\bibinfo  {journal} {arXiv:1506.08479}\ } (\bibinfo {year}
  {2015}{\natexlab{a}})}\BibitemShut {NoStop}%
\bibitem [{\citenamefont {Rieffel}\ \emph {et~al.}(2015)\citenamefont
  {Rieffel}, \citenamefont {Venturelli}, \citenamefont {O'Gorman},
  \citenamefont {Do}, \citenamefont {Prystay},\ and\ \citenamefont
  {Smelyanskiy}}]{Rieffel:2015aa}%
  \BibitemOpen
  \bibfield  {author} {\bibinfo {author} {\bibfnamefont {E.~G.}\ \bibnamefont
  {Rieffel}}, \bibinfo {author} {\bibfnamefont {D.}~\bibnamefont {Venturelli}},
  \bibinfo {author} {\bibfnamefont {B.}~\bibnamefont {O'Gorman}}, \bibinfo
  {author} {\bibfnamefont {M.~B.}\ \bibnamefont {Do}}, \bibinfo {author}
  {\bibfnamefont {E.~M.}\ \bibnamefont {Prystay}}, \ and\ \bibinfo {author}
  {\bibfnamefont {V.~N.}\ \bibnamefont {Smelyanskiy}},\ }\href {\doibase
  10.1007/s11128-014-0892-x} {\bibfield  {journal} {\bibinfo  {journal}
  {Quantum Information Processing}\ }\textbf {\bibinfo {volume} {14}},\
  \bibinfo {pages} {1} (\bibinfo {year} {2015})}\BibitemShut {NoStop}%
\bibitem [{\citenamefont {Rosenberg}\ \emph {et~al.}(2015)\citenamefont
  {Rosenberg}, \citenamefont {Haghnegahdar}, \citenamefont {Goddard},
  \citenamefont {Carr}, \citenamefont {Wu},\ and\ \citenamefont
  {Prado}}]{Rosenberg:2015}%
  \BibitemOpen
  \bibfield  {author} {\bibinfo {author} {\bibfnamefont {G.}~\bibnamefont
  {Rosenberg}}, \bibinfo {author} {\bibfnamefont {P.}~\bibnamefont
  {Haghnegahdar}}, \bibinfo {author} {\bibfnamefont {P.}~\bibnamefont
  {Goddard}}, \bibinfo {author} {\bibfnamefont {P.}~\bibnamefont {Carr}},
  \bibinfo {author} {\bibfnamefont {K.}~\bibnamefont {Wu}}, \ and\ \bibinfo
  {author} {\bibfnamefont {M.~L.~d.}\ \bibnamefont {Prado}},\ }\href
  {http://arXiv.org/abs/1508.06182} {\bibfield  {journal} {\bibinfo  {journal}
  {arXiv:1508.06182}\ } (\bibinfo {year} {2015})}\BibitemShut {NoStop}%
\bibitem [{\citenamefont {Lucas}(2014)}]{2013arXiv1302.5843L}%
  \BibitemOpen
  \bibfield  {author} {\bibinfo {author} {\bibfnamefont {A.}~\bibnamefont
  {Lucas}},\ }\href {\doibase 10.3389/fphy.2014.00005} {\bibfield  {journal}
  {\bibinfo  {journal} {Front. Phys.}\ }\textbf {\bibinfo {volume} {2}},\
  \bibinfo {pages} {5} (\bibinfo {year} {2014})}\BibitemShut {NoStop}%
\bibitem [{\citenamefont {Choi}(2008)}]{Choi1}%
  \BibitemOpen
  \bibfield  {author} {\bibinfo {author} {\bibfnamefont {V.}~\bibnamefont
  {Choi}},\ }\href {\doibase 10.1007/s11128-008-0082-9} {\bibfield  {journal}
  {\bibinfo  {journal} {Quant. Inf. Proc.}\ }\textbf {\bibinfo {volume} {7}},\
  \bibinfo {pages} {193} (\bibinfo {year} {2008})}\BibitemShut {NoStop}%
\bibitem [{\citenamefont {Choi}(2011)}]{Choi2}%
  \BibitemOpen
  \bibfield  {author} {\bibinfo {author} {\bibfnamefont {V.}~\bibnamefont
  {Choi}},\ }\href {\doibase 10.1007/s11128-010-0200-3} {\bibfield  {journal}
  {\bibinfo  {journal} {Quant. Inf. Proc.}\ }\textbf {\bibinfo {volume} {10}},\
  \bibinfo {pages} {343} (\bibinfo {year} {2011})}\BibitemShut {NoStop}%
\bibitem [{\citenamefont {Klymko}\ \emph {et~al.}(2014)\citenamefont {Klymko},
  \citenamefont {Sullivan},\ and\ \citenamefont
  {Humble}}]{klymko_adiabatic_2012}%
  \BibitemOpen
  \bibfield  {author} {\bibinfo {author} {\bibfnamefont {C.}~\bibnamefont
  {Klymko}}, \bibinfo {author} {\bibfnamefont {B.~D.}\ \bibnamefont
  {Sullivan}}, \ and\ \bibinfo {author} {\bibfnamefont {T.~S.}\ \bibnamefont
  {Humble}},\ }\href {\doibase 10.1007/s11128-013-0683-9} {\bibfield  {journal}
  {\bibinfo  {journal} {Quant. Inf. Proc.}\ }\textbf {\bibinfo {volume} {13}},\
  \bibinfo {pages} {709} (\bibinfo {year} {2014})}\BibitemShut {NoStop}%
\bibitem [{\citenamefont {Venturelli}\ \emph
  {et~al.}(2015{\natexlab{b}})\citenamefont {Venturelli}, \citenamefont
  {Mandr{\`a}}, \citenamefont {Knysh}, \citenamefont {O'Gorman}, \citenamefont
  {Biswas},\ and\ \citenamefont {Smelyanskiy}}]{Venturelli:2014nx}%
  \BibitemOpen
  \bibfield  {author} {\bibinfo {author} {\bibfnamefont {D.}~\bibnamefont
  {Venturelli}}, \bibinfo {author} {\bibfnamefont {S.}~\bibnamefont
  {Mandr{\`a}}}, \bibinfo {author} {\bibfnamefont {S.}~\bibnamefont {Knysh}},
  \bibinfo {author} {\bibfnamefont {B.}~\bibnamefont {O'Gorman}}, \bibinfo
  {author} {\bibfnamefont {R.}~\bibnamefont {Biswas}}, \ and\ \bibinfo {author}
  {\bibfnamefont {V.}~\bibnamefont {Smelyanskiy}},\ }\href
  {http://link.aps.org/doi/10.1103/PhysRevX.5.031040} {\bibfield  {journal}
  {\bibinfo  {journal} {Phys. Rev. X}\ }\textbf {\bibinfo {volume} {5}},\
  \bibinfo {pages} {031040} (\bibinfo {year} {2015}{\natexlab{b}})}\BibitemShut
  {NoStop}%
\bibitem [{\citenamefont {Rubin}\ \emph {et~al.}(1981)\citenamefont {Rubin}
  \emph {et~al.}}]{rubin1981bayesian}%
  \BibitemOpen
  \bibfield  {author} {\bibinfo {author} {\bibfnamefont {D.~B.}\ \bibnamefont
  {Rubin}} \emph {et~al.},\ }\href@noop {} {\bibfield  {journal} {\bibinfo
  {journal} {The annals of statistics}\ }\textbf {\bibinfo {volume} {9}},\
  \bibinfo {pages} {130} (\bibinfo {year} {1981})}\BibitemShut {NoStop}%
\bibitem [{\citenamefont {Albash}\ and\ \citenamefont
  {Lidar}(2017)}]{Albash:2017aa}%
  \BibitemOpen
  \bibfield  {author} {\bibinfo {author} {\bibfnamefont {T.}~\bibnamefont
  {Albash}}\ and\ \bibinfo {author} {\bibfnamefont {D.~A.}\ \bibnamefont
  {Lidar}},\ }\href {http://arXiv.org/abs/1705.07452} {\bibfield  {journal}
  {\bibinfo  {journal} {arXiv:1705.07452}\ } (\bibinfo {year}
  {2017})}\BibitemShut {NoStop}%
\bibitem [{\citenamefont {Katzgraber}\ \emph {et~al.}(2014)\citenamefont
  {Katzgraber}, \citenamefont {Hamze},\ and\ \citenamefont
  {Andrist}}]{2014Katzgraber}%
  \BibitemOpen
  \bibfield  {author} {\bibinfo {author} {\bibfnamefont {H.~G.}\ \bibnamefont
  {Katzgraber}}, \bibinfo {author} {\bibfnamefont {F.}~\bibnamefont {Hamze}}, \
  and\ \bibinfo {author} {\bibfnamefont {R.~S.}\ \bibnamefont {Andrist}},\
  }\href {http://link.aps.org/doi/10.1103/PhysRevX.4.021008} {\bibfield
  {journal} {\bibinfo  {journal} {Phys. Rev. X}\ }\textbf {\bibinfo {volume}
  {4}},\ \bibinfo {pages} {021008} (\bibinfo {year} {2014})}\BibitemShut
  {NoStop}%
\bibitem [{\citenamefont {Harrow}\ \emph {et~al.}(2009)\citenamefont {Harrow},
  \citenamefont {Hassidim},\ and\ \citenamefont
  {Lloyd}}]{PhysRevLett.103.150502}%
  \BibitemOpen
  \bibfield  {author} {\bibinfo {author} {\bibfnamefont {A.~W.}\ \bibnamefont
  {Harrow}}, \bibinfo {author} {\bibfnamefont {A.}~\bibnamefont {Hassidim}}, \
  and\ \bibinfo {author} {\bibfnamefont {S.}~\bibnamefont {Lloyd}},\
  }\href@noop {} {\bibfield  {journal} {\bibinfo  {journal} {Phys. Rev. Lett.}\
  }\textbf {\bibinfo {volume} {103}},\ \bibinfo {pages} {150502} (\bibinfo
  {year} {2009})}\BibitemShut {NoStop}%
\bibitem [{\citenamefont {Hen}\ \emph {et~al.}(2015)\citenamefont {Hen},
  \citenamefont {Job}, \citenamefont {Albash}, \citenamefont {R{\o}nnow},
  \citenamefont {Troyer},\ and\ \citenamefont {Lidar}}]{Hen:2015rt}%
  \BibitemOpen
  \bibfield  {author} {\bibinfo {author} {\bibfnamefont {I.}~\bibnamefont
  {Hen}}, \bibinfo {author} {\bibfnamefont {J.}~\bibnamefont {Job}}, \bibinfo
  {author} {\bibfnamefont {T.}~\bibnamefont {Albash}}, \bibinfo {author}
  {\bibfnamefont {T.~F.}\ \bibnamefont {R{\o}nnow}}, \bibinfo {author}
  {\bibfnamefont {M.}~\bibnamefont {Troyer}}, \ and\ \bibinfo {author}
  {\bibfnamefont {D.~A.}\ \bibnamefont {Lidar}},\ }\href
  {http://link.aps.org/doi/10.1103/PhysRevA.92.042325} {\bibfield  {journal}
  {\bibinfo  {journal} {{Phys. Rev. A}}\ }\textbf {\bibinfo {volume} {92}},\
  \bibinfo {pages} {042325} (\bibinfo {year} {2015})}\BibitemShut {NoStop}%
\bibitem [{\citenamefont {King}\ \emph
  {et~al.}(2015{\natexlab{b}})\citenamefont {King}, \citenamefont {Lanting},\
  and\ \citenamefont {Harris}}]{King:2015zr}%
  \BibitemOpen
  \bibfield  {author} {\bibinfo {author} {\bibfnamefont {A.~D.}\ \bibnamefont
  {King}}, \bibinfo {author} {\bibfnamefont {T.}~\bibnamefont {Lanting}}, \
  and\ \bibinfo {author} {\bibfnamefont {R.}~\bibnamefont {Harris}},\ }\href
  {http://arXiv.org/abs/1502.02098} {\bibfield  {journal} {\bibinfo  {journal}
  {arXiv:1502.02098}\ } (\bibinfo {year} {2015}{\natexlab{b}})}\BibitemShut
  {NoStop}%
\bibitem [{\citenamefont {Selby}(2013)}]{selby:13b}%
  \BibitemOpen
  \bibfield  {author} {\bibinfo {author} {\bibfnamefont {A.}~\bibnamefont
  {Selby}},\ }\href@noop {} {\enquote {\bibinfo {title} {D-wave: comment on
  comparison with classical computers},}\ }\bibinfo {howpublished}
  {\url{http://tinyurl.com/dwave-vs-classical}} (\bibinfo {year}
  {2013})\BibitemShut {NoStop}%
\bibitem [{\citenamefont {Ferguson}(2008)}]{Ferguson:book}%
  \BibitemOpen
  \bibfield  {author} {\bibinfo {author} {\bibfnamefont {T.~S.}\ \bibnamefont
  {Ferguson}},\ }\href {http://www.e-booksdirectory.com/details.php?ebook=5651}
  {\enquote {\bibinfo {title} {{Optimal Stopping and Applications}},}\ }
  (\bibinfo {year} {2008})\BibitemShut {NoStop}%
\bibitem [{\citenamefont {King}\ \emph {et~al.}(2017)\citenamefont {King},
  \citenamefont {Yarkoni}, \citenamefont {Raymond}, \citenamefont {Ozfidan},
  \citenamefont {King}, \citenamefont {Nevisi}, \citenamefont {Hilton},\ and\
  \citenamefont {McGeoch}}]{DW2000Q}%
  \BibitemOpen
  \bibfield  {author} {\bibinfo {author} {\bibfnamefont {J.}~\bibnamefont
  {King}}, \bibinfo {author} {\bibfnamefont {S.}~\bibnamefont {Yarkoni}},
  \bibinfo {author} {\bibfnamefont {J.}~\bibnamefont {Raymond}}, \bibinfo
  {author} {\bibfnamefont {I.}~\bibnamefont {Ozfidan}}, \bibinfo {author}
  {\bibfnamefont {A.~D.}\ \bibnamefont {King}}, \bibinfo {author}
  {\bibfnamefont {M.~M.}\ \bibnamefont {Nevisi}}, \bibinfo {author}
  {\bibfnamefont {J.~P.}\ \bibnamefont {Hilton}}, \ and\ \bibinfo {author}
  {\bibfnamefont {C.~C.}\ \bibnamefont {McGeoch}},\ }\href
  {http://arXiv.org/abs/1701.04579} {\bibfield  {journal} {\bibinfo  {journal}
  {arXiv:1701.04579}\ } (\bibinfo {year} {2017})}\BibitemShut {NoStop}%
\bibitem [{\citenamefont {Roland}\ and\ \citenamefont
  {Cerf}(2002)}]{Roland:2002ul}%
  \BibitemOpen
  \bibfield  {author} {\bibinfo {author} {\bibfnamefont {J.}~\bibnamefont
  {Roland}}\ and\ \bibinfo {author} {\bibfnamefont {N.~J.}\ \bibnamefont
  {Cerf}},\ }\href {http://link.aps.org/doi/10.1103/PhysRevA.65.042308}
  {\bibfield  {journal} {\bibinfo  {journal} {Phys. Rev. A}\ }\textbf {\bibinfo
  {volume} {65}},\ \bibinfo {pages} {042308} (\bibinfo {year}
  {2002})}\BibitemShut {NoStop}%
\bibitem [{\citenamefont {Rezakhani}\ \emph {et~al.}(2010)\citenamefont
  {Rezakhani}, \citenamefont {Pimachev},\ and\ \citenamefont {Lidar}}]{RPL:10}%
  \BibitemOpen
  \bibfield  {author} {\bibinfo {author} {\bibfnamefont {A.~T.}\ \bibnamefont
  {Rezakhani}}, \bibinfo {author} {\bibfnamefont {A.~K.}\ \bibnamefont
  {Pimachev}}, \ and\ \bibinfo {author} {\bibfnamefont {D.~A.}\ \bibnamefont
  {Lidar}},\ }\href {http://link.aps.org/doi/10.1103/PhysRevA.82.052305}
  {\bibfield  {journal} {\bibinfo  {journal} {Phys. Rev. A}\ }\textbf {\bibinfo
  {volume} {82}},\ \bibinfo {pages} {052305} (\bibinfo {year}
  {2010})}\BibitemShut {NoStop}%
\bibitem [{\citenamefont {Pudenz}\ \emph {et~al.}(2014)\citenamefont {Pudenz},
  \citenamefont {Albash},\ and\ \citenamefont {Lidar}}]{PAL:13}%
  \BibitemOpen
  \bibfield  {author} {\bibinfo {author} {\bibfnamefont {K.~L.}\ \bibnamefont
  {Pudenz}}, \bibinfo {author} {\bibfnamefont {T.}~\bibnamefont {Albash}}, \
  and\ \bibinfo {author} {\bibfnamefont {D.~A.}\ \bibnamefont {Lidar}},\ }\href
  {\doibase 10.1038/ncomms4243} {\bibfield  {journal} {\bibinfo  {journal}
  {Nat. Commun.}\ }\textbf {\bibinfo {volume} {5}},\ \bibinfo {pages} {3243}
  (\bibinfo {year} {2014})}\BibitemShut {NoStop}%
\bibitem [{\citenamefont {Pudenz}\ \emph {et~al.}(2015)\citenamefont {Pudenz},
  \citenamefont {Albash},\ and\ \citenamefont {Lidar}}]{PAL:14}%
  \BibitemOpen
  \bibfield  {author} {\bibinfo {author} {\bibfnamefont {K.~L.}\ \bibnamefont
  {Pudenz}}, \bibinfo {author} {\bibfnamefont {T.}~\bibnamefont {Albash}}, \
  and\ \bibinfo {author} {\bibfnamefont {D.~A.}\ \bibnamefont {Lidar}},\ }\href
  {http://link.aps.org/doi/10.1103/PhysRevA.91.042302} {\bibfield  {journal}
  {\bibinfo  {journal} {{Phys. Rev. A}}\ }\textbf {\bibinfo {volume} {91}},\
  \bibinfo {pages} {042302} (\bibinfo {year} {2015})}\BibitemShut {NoStop}%
\bibitem [{\citenamefont {Vinci}\ \emph {et~al.}(2015)\citenamefont {Vinci},
  \citenamefont {Albash}, \citenamefont {Paz-Silva}, \citenamefont {Hen},\ and\
  \citenamefont {Lidar}}]{Vinci:2015jt}%
  \BibitemOpen
  \bibfield  {author} {\bibinfo {author} {\bibfnamefont {W.}~\bibnamefont
  {Vinci}}, \bibinfo {author} {\bibfnamefont {T.}~\bibnamefont {Albash}},
  \bibinfo {author} {\bibfnamefont {G.}~\bibnamefont {Paz-Silva}}, \bibinfo
  {author} {\bibfnamefont {I.}~\bibnamefont {Hen}}, \ and\ \bibinfo {author}
  {\bibfnamefont {D.~A.}\ \bibnamefont {Lidar}},\ }\href
  {http://link.aps.org/doi/10.1103/PhysRevA.92.042310} {\bibfield  {journal}
  {\bibinfo  {journal} {{Phys. Rev. A}}\ }\textbf {\bibinfo {volume} {92}},\
  \bibinfo {pages} {042310} (\bibinfo {year} {2015})}\BibitemShut {NoStop}%
\bibitem [{\citenamefont {Barahona}(1982)}]{Barahona1982}%
  \BibitemOpen
  \bibfield  {author} {\bibinfo {author} {\bibfnamefont {F.}~\bibnamefont
  {Barahona}},\ }\href {http://stacks.iop.org/0305-4470/15/i=10/a=028}
  {\bibfield  {journal} {\bibinfo  {journal} {J. Phys. A: Math. Gen}\ }\textbf
  {\bibinfo {volume} {15}},\ \bibinfo {pages} {3241} (\bibinfo {year}
  {1982})}\BibitemShut {NoStop}%
\bibitem [{\citenamefont {Vinci}\ \emph {et~al.}(2016)\citenamefont {Vinci},
  \citenamefont {Albash},\ and\ \citenamefont {Lidar}}]{vinci2015nested}%
  \BibitemOpen
  \bibfield  {author} {\bibinfo {author} {\bibfnamefont {W.}~\bibnamefont
  {Vinci}}, \bibinfo {author} {\bibfnamefont {T.}~\bibnamefont {Albash}}, \
  and\ \bibinfo {author} {\bibfnamefont {D.~A.}\ \bibnamefont {Lidar}},\ }\href
  {http://dx.doi.org/10.1038/npjqi.2016.17} {\bibfield  {journal} {\bibinfo
  {journal} {Nature Quantum Information}\ }\textbf {\bibinfo {volume} {2}},\
  \bibinfo {pages} {16017} (\bibinfo {year} {2016})}\BibitemShut {NoStop}%
\bibitem [{\citenamefont {Aliferis}\ \emph {et~al.}(2006)\citenamefont
  {Aliferis}, \citenamefont {Gottesman},\ and\ \citenamefont
  {Preskill}}]{Aliferis:05}%
  \BibitemOpen
  \bibfield  {author} {\bibinfo {author} {\bibfnamefont {P.}~\bibnamefont
  {Aliferis}}, \bibinfo {author} {\bibfnamefont {D.}~\bibnamefont {Gottesman}},
  \ and\ \bibinfo {author} {\bibfnamefont {J.}~\bibnamefont {Preskill}},\
  }\href {http://www.rintonpress.com/xqic6/qic-6-2/097-165.pdf} {\bibfield
  {journal} {\bibinfo  {journal} {Quantum Inf. Comput.}\ }\textbf {\bibinfo
  {volume} {6}},\ \bibinfo {pages} {97} (\bibinfo {year} {2006})}\BibitemShut
  {NoStop}%
\bibitem [{\citenamefont {Raussendorf}(2012)}]{Raussendorf28092012}%
  \BibitemOpen
  \bibfield  {author} {\bibinfo {author} {\bibfnamefont {R.}~\bibnamefont
  {Raussendorf}},\ }\href {\doibase 10.1098/rsta.2011.0494} {\bibfield
  {journal} {\bibinfo  {journal} {Philosophical Transactions of the Royal
  Society A: Mathematical, Physical and Engineering Sciences}\ }\textbf
  {\bibinfo {volume} {370}},\ \bibinfo {pages} {4541} (\bibinfo {year}
  {2012})}\BibitemShut {NoStop}%
\bibitem [{\citenamefont {Lidar}\ and\ \citenamefont
  {Brun}(2013)}]{Lidar-Brun:book}%
  \BibitemOpen
  \bibinfo {editor} {\bibfnamefont {D.}~\bibnamefont {Lidar}}\ and\ \bibinfo
  {editor} {\bibfnamefont {T.}~\bibnamefont {Brun}},\ eds.,\ \href
  {http://www.cambridge.org/9780521897877} {\emph {\bibinfo {title} {Quantum
  Error Correction}}}\ (\bibinfo  {publisher} {Cambridge University Press},\
  \bibinfo {address} {{Cambridge, UK}},\ \bibinfo {year} {2013})\BibitemShut
  {NoStop}%
\bibitem [{\citenamefont {Jordan}\ \emph {et~al.}(2006)\citenamefont {Jordan},
  \citenamefont {Farhi},\ and\ \citenamefont {Shor}}]{jordan2006error}%
  \BibitemOpen
  \bibfield  {author} {\bibinfo {author} {\bibfnamefont {S.~P.}\ \bibnamefont
  {Jordan}}, \bibinfo {author} {\bibfnamefont {E.}~\bibnamefont {Farhi}}, \
  and\ \bibinfo {author} {\bibfnamefont {P.~W.}\ \bibnamefont {Shor}},\ }\href
  {http://link.aps.org/doi/10.1103/PhysRevA.74.052322} {\bibfield  {journal}
  {\bibinfo  {journal} {{Phys. Rev. A}}\ }\textbf {\bibinfo {volume} {74}},\
  \bibinfo {pages} {052322} (\bibinfo {year} {2006})}\BibitemShut {NoStop}%
\bibitem [{\citenamefont {Lidar}(2008)}]{PhysRevLett.100.160506}%
  \BibitemOpen
  \bibfield  {author} {\bibinfo {author} {\bibfnamefont {D.~A.}\ \bibnamefont
  {Lidar}},\ }\href {http://link.aps.org/doi/10.1103/PhysRevLett.100.160506}
  {\bibfield  {journal} {\bibinfo  {journal} {{Phys.~Rev.~Lett.}}\ }\textbf
  {\bibinfo {volume} {100}},\ \bibinfo {pages} {160506} (\bibinfo {year}
  {2008})}\BibitemShut {NoStop}%
\bibitem [{\citenamefont {Paz-Silva}\ \emph {et~al.}(2012)\citenamefont
  {Paz-Silva}, \citenamefont {Rezakhani}, \citenamefont {Dominy},\ and\
  \citenamefont {Lidar}}]{PhysRevLett.108.080501}%
  \BibitemOpen
  \bibfield  {author} {\bibinfo {author} {\bibfnamefont {G.~A.}\ \bibnamefont
  {Paz-Silva}}, \bibinfo {author} {\bibfnamefont {A.~T.}\ \bibnamefont
  {Rezakhani}}, \bibinfo {author} {\bibfnamefont {J.~M.}\ \bibnamefont
  {Dominy}}, \ and\ \bibinfo {author} {\bibfnamefont {D.~A.}\ \bibnamefont
  {Lidar}},\ }\href {\doibase 10.1103/PhysRevLett.108.080501} {\bibfield
  {journal} {\bibinfo  {journal} {Phys. Rev. Lett.}\ }\textbf {\bibinfo
  {volume} {108}},\ \bibinfo {pages} {080501} (\bibinfo {year}
  {2012})}\BibitemShut {NoStop}%
\bibitem [{\citenamefont {Young}\ \emph {et~al.}(2013)\citenamefont {Young},
  \citenamefont {Sarovar},\ and\ \citenamefont {Blume-Kohout}}]{Young:13}%
  \BibitemOpen
  \bibfield  {author} {\bibinfo {author} {\bibfnamefont {K.~C.}\ \bibnamefont
  {Young}}, \bibinfo {author} {\bibfnamefont {M.}~\bibnamefont {Sarovar}}, \
  and\ \bibinfo {author} {\bibfnamefont {R.}~\bibnamefont {Blume-Kohout}},\
  }\href {http://link.aps.org/doi/10.1103/PhysRevX.3.041013} {\bibfield
  {journal} {\bibinfo  {journal} {Phys. Rev. X}\ }\textbf {\bibinfo {volume}
  {3}},\ \bibinfo {pages} {041013} (\bibinfo {year} {2013})}\BibitemShut
  {NoStop}%
\bibitem [{\citenamefont {Matsuura}\ \emph {et~al.}(2016)\citenamefont
  {Matsuura}, \citenamefont {Nishimori}, \citenamefont {Albash},\ and\
  \citenamefont {Lidar}}]{MNAL:15}%
  \BibitemOpen
  \bibfield  {author} {\bibinfo {author} {\bibfnamefont {S.}~\bibnamefont
  {Matsuura}}, \bibinfo {author} {\bibfnamefont {H.}~\bibnamefont {Nishimori}},
  \bibinfo {author} {\bibfnamefont {T.}~\bibnamefont {Albash}}, \ and\ \bibinfo
  {author} {\bibfnamefont {D.~A.}\ \bibnamefont {Lidar}},\ }\href
  {http://link.aps.org/doi/10.1103/PhysRevLett.116.220501} {\bibfield
  {journal} {\bibinfo  {journal} {Physical Review Letters}\ }\textbf {\bibinfo
  {volume} {116}},\ \bibinfo {pages} {220501} (\bibinfo {year}
  {2016})}\BibitemShut {NoStop}%
\bibitem [{\citenamefont {Matsuura}\ \emph {et~al.}(2017)\citenamefont
  {Matsuura}, \citenamefont {Nishimori}, \citenamefont {Vinci}, \citenamefont
  {Albash},\ and\ \citenamefont {Lidar}}]{matsuura2017qac}%
  \BibitemOpen
  \bibfield  {author} {\bibinfo {author} {\bibfnamefont {S.}~\bibnamefont
  {Matsuura}}, \bibinfo {author} {\bibfnamefont {H.}~\bibnamefont {Nishimori}},
  \bibinfo {author} {\bibfnamefont {W.}~\bibnamefont {Vinci}}, \bibinfo
  {author} {\bibfnamefont {T.}~\bibnamefont {Albash}}, \ and\ \bibinfo {author}
  {\bibfnamefont {D.~A.}\ \bibnamefont {Lidar}},\ }\href {\doibase
  10.1103/PhysRevA.95.022308} {\bibfield  {journal} {\bibinfo  {journal} {Phys.
  Rev. A}\ }\textbf {\bibinfo {volume} {95}},\ \bibinfo {pages} {022308}
  (\bibinfo {year} {2017})}\BibitemShut {NoStop}%
\bibitem [{\citenamefont {Mishra}\ \emph {et~al.}(2015)\citenamefont {Mishra},
  \citenamefont {Albash},\ and\ \citenamefont {Lidar}}]{Mishra:2015}%
  \BibitemOpen
  \bibfield  {author} {\bibinfo {author} {\bibfnamefont {A.}~\bibnamefont
  {Mishra}}, \bibinfo {author} {\bibfnamefont {T.}~\bibnamefont {Albash}}, \
  and\ \bibinfo {author} {\bibfnamefont {D.~A.}\ \bibnamefont {Lidar}},\ }\href
  {\doibase 10.1007/s11128-015-1201-z} {\bibfield  {journal} {\bibinfo
  {journal} {Quant. Inf. Proc.}\ }\textbf {\bibinfo {volume} {15}},\ \bibinfo
  {pages} {609} (\bibinfo {year} {2015})}\BibitemShut {NoStop}%
\bibitem [{\citenamefont {Marvian}\ and\ \citenamefont
  {Lidar}(2017)}]{Marvian-Lidar:16}%
  \BibitemOpen
  \bibfield  {author} {\bibinfo {author} {\bibfnamefont {M.}~\bibnamefont
  {Marvian}}\ and\ \bibinfo {author} {\bibfnamefont {D.~A.}\ \bibnamefont
  {Lidar}},\ }\href {https://link.aps.org/doi/10.1103/PhysRevLett.118.030504}
  {\bibfield  {journal} {\bibinfo  {journal} {Physical Review Letters}\
  }\textbf {\bibinfo {volume} {118}},\ \bibinfo {pages} {030504} (\bibinfo
  {year} {2017})}\BibitemShut {NoStop}%
\bibitem [{\citenamefont {Jiang}\ and\ \citenamefont
  {Rieffel}(2017)}]{Jiang:2015kx}%
  \BibitemOpen
  \bibfield  {author} {\bibinfo {author} {\bibfnamefont {Z.}~\bibnamefont
  {Jiang}}\ and\ \bibinfo {author} {\bibfnamefont {E.~G.}\ \bibnamefont
  {Rieffel}},\ }\href {\doibase 10.1007/s11128-017-1527-9} {\bibfield
  {journal} {\bibinfo  {journal} {Quantum Information Processing}\ }\textbf
  {\bibinfo {volume} {16}},\ \bibinfo {pages} {89} (\bibinfo {year}
  {2017})}\BibitemShut {NoStop}%
\bibitem [{\citenamefont {Inagaki}\ \emph {et~al.}(2016)\citenamefont
  {Inagaki}, \citenamefont {Haribara}, \citenamefont {Igarashi}, \citenamefont
  {Sonobe}, \citenamefont {Tamate}, \citenamefont {Honjo}, \citenamefont
  {Marandi}, \citenamefont {McMahon}, \citenamefont {Umeki}, \citenamefont
  {Enbutsu}, \citenamefont {Tadanaga}, \citenamefont {Takenouchi},
  \citenamefont {Aihara}, \citenamefont {Kawarabayashi}, \citenamefont {Inoue},
  \citenamefont {Utsunomiya},\ and\ \citenamefont {Takesue}}]{Inagaki:2016aa}%
  \BibitemOpen
  \bibfield  {author} {\bibinfo {author} {\bibfnamefont {T.}~\bibnamefont
  {Inagaki}}, \bibinfo {author} {\bibfnamefont {Y.}~\bibnamefont {Haribara}},
  \bibinfo {author} {\bibfnamefont {K.}~\bibnamefont {Igarashi}}, \bibinfo
  {author} {\bibfnamefont {T.}~\bibnamefont {Sonobe}}, \bibinfo {author}
  {\bibfnamefont {S.}~\bibnamefont {Tamate}}, \bibinfo {author} {\bibfnamefont
  {T.}~\bibnamefont {Honjo}}, \bibinfo {author} {\bibfnamefont
  {A.}~\bibnamefont {Marandi}}, \bibinfo {author} {\bibfnamefont {P.~L.}\
  \bibnamefont {McMahon}}, \bibinfo {author} {\bibfnamefont {T.}~\bibnamefont
  {Umeki}}, \bibinfo {author} {\bibfnamefont {K.}~\bibnamefont {Enbutsu}},
  \bibinfo {author} {\bibfnamefont {O.}~\bibnamefont {Tadanaga}}, \bibinfo
  {author} {\bibfnamefont {H.}~\bibnamefont {Takenouchi}}, \bibinfo {author}
  {\bibfnamefont {K.}~\bibnamefont {Aihara}}, \bibinfo {author} {\bibfnamefont
  {K.-i.}\ \bibnamefont {Kawarabayashi}}, \bibinfo {author} {\bibfnamefont
  {K.}~\bibnamefont {Inoue}}, \bibinfo {author} {\bibfnamefont
  {S.}~\bibnamefont {Utsunomiya}}, \ and\ \bibinfo {author} {\bibfnamefont
  {H.}~\bibnamefont {Takesue}},\ }\href
  {http://science.sciencemag.org/content/354/6312/603.abstract} {\bibfield
  {journal} {\bibinfo  {journal} {Science}\ }\textbf {\bibinfo {volume}
  {354}},\ \bibinfo {pages} {603} (\bibinfo {year} {2016})}\BibitemShut
  {NoStop}%
\end{thebibliography}
\end{document}